\documentclass[12pt,a4paper]{article}
\usepackage{graphicx}
\usepackage[T1]{fontenc}
\usepackage[utf8]{inputenc}
\usepackage{textcomp}
\usepackage[sc,osf]{mathpazo}
\usepackage{a4wide}  
\usepackage{latexsym,amsthm,amsfonts,amsmath,mathrsfs,amssymb}
\usepackage{dsfont}
\usepackage{accents}
\usepackage[nosort]{cite}
\usepackage{booktabs} 
\usepackage[unicode,implicit]{hyperref}
\hypersetup{%
  pdftitle    = {The first law and Wald entropy formula of heterotic stringy
    black holes at first order in $\alpha'$}
  pdfkeywords = {duality, T duality, strings, heterotic superstrings, effective action, gauge symmetry, entropy,  black holes, Noether, Wald, compactification, momentum maps, thermodynamics, corrections, higher order, alpha prime}
  pdfauthor   = {Zachary Elgood, Tom\'as Ort\'{\i}n  and David Pere\~n\'{\i}guez},
  plainpages  = true,
  colorlinks  = true,
  citecolor   = blue,
  urlcolor    = red,
  linkcolor   = black
}
\newcommand{\hepth}[1]{{\tt
\href{http://www.arXiv.org/abs/hep-th/#1}{hep-th/#1}}}
\newcommand{\grqc}[1]{{\tt
\href{http://www.arXiv.org/abs/gr-qc/#1}{gr-qc/#1}}}

\newcommand{\arxiv}[1]{{\tt arXiv:\href{http://www.arXiv.org/abs/#1}{#1}}}
\newcommand{\doi}[1]{{\tt DOI:\href{http://dx.doi.org/#1}{#1}}}

\makeatletter
\@addtoreset{equation}{section}
\makeatother

\begin{document}

\begin{flushright}
  \small
IFT-UAM/CSIC-20-182\\
December 29\textsuperscript{th}, 2020\\
\normalsize
\end{flushright}

\vspace{1cm}

\begin{center}

  {\Large {\bf The first law and Wald entropy formula\\[.5cm]
      of heterotic stringy black holes at first order in $\alpha'$}}

\vspace{1.5cm}

\renewcommand{\thefootnote}{\alph{footnote}}

{\sl\large Zachary Elgood,}\footnote{Email: {\tt zachary.elgood[at]uam.es}}
{\sl\large Tom\'{a}s Ort\'{\i}n}\footnote{Email: {\tt tomas.ortin[at]csic.es}}
{\sl\large and David Pere\~n\'{\i}guez,}\footnote{Email: {\tt david.perenniguez[at]uam.es}}

\setcounter{footnote}{0}
\renewcommand{\thefootnote}{\arabic{footnote}}

\vspace{1cm}

{\it Instituto de F\'{\i}sica Te\'orica UAM/CSIC\\
C/ Nicol\'as Cabrera, 13--15,  C.U.~Cantoblanco, E-28049 Madrid, Spain}

\vspace{1cm}


{\bf Abstract}
\end{center}
\begin{quotation}
  {\small We derive the first law of black hole mechanics in the
    context of the Heterotic Superstring effective action to first
    order in $\alpha'$ using Wald's formalism. We carefully take into
    account all the symmetries of the theory and, as a result, we
    obtain a manifestly gauge- and Lorentz-invariant entropy formula
    in which all the terms can be computed explicitly. An entropy
    formula with these properties allows unambiguous calculations of
    macroscopic black-hole entropies to first order in $\alpha'$ that
    can be reliably used in a comparison with the microscopic
    ones. Such a formula was still lacking in the literature.

    In the proof we use momentum maps to define covariant variations and Lie
    derivatives and \textit{restricted generalized zeroth laws} which state
    the closedness of certain differential forms on the bifurcation sphere and
    imply the constancy of the associated potentials on it.

    We study the relation between our entropy formula and other formulae that
    have been used in the literature.
  }
\end{quotation}

\newpage
\pagestyle{plain}

\tableofcontents


\section{Introduction}

The interpretation of the black-hole entropy in terms of the
degeneracy of string microstates is, beyond any doubt, one of the main
achievements of String Theory \cite{Strominger:1996sh}. This
interpretation relies, on the one hand, on the correct identification
of the black-hole charges in terms of branes whose presence affects
the quantization of the string. On the other, it depends on a correct
calculation of the macroscopic entropy. In simple cases, at leading
order in $\alpha'$, the identification of the field fluxes with the
brane sources that would produce them is straightforward and, also,
the macroscopic entropy is given by the Bekenstein-Hawking formula
$S=\mathcal{A}_{\mathcal{H}}/(4G_{N})$, where
$\mathcal{A}_{\mathcal{H}}$ is the area of the horizon. In more
complicated cases, the couplings can make the identification of the
brane sources through the charges more complicated
\cite{Faedo:2019xii} and, beyond leading order in $\alpha'$, the
presence of terms of higher order in the curvature and, in the
Heterotic Superstring case, of complicated Yang-Mills (YM) and Lorentz
Chern-Simons terms \cite{Bergshoeff:1989de} can also make the
calculation of the macroscopic entropy very difficult. This is the
problem we will deal with in this paper.

The standard method to calculate the black-hole entropy in theories of
higher order in the curvature is to use Wald's formalism
\cite{Lee:1990nz,Wald:1993nt}, usually applying directly the Iyer-Wald
prescription \cite{Iyer:1994ys}. As we have recently discussed in
Refs.~\cite{Elgood:2020xwu,Elgood:2020svt,Ortin:2020xdm} (see also
references therein), the Iyer-Wald prescription was derived assuming
that all the fields of the theory behave as tensors under
diffeomorphisms which, as matter of fact, is only true for the metric
and uncharged scalars. All the fields of the Standard Model, except
for the metric, have some kind of gauge freedom and do not transform
as tensors under diffeomorphisms. Even the gravitational field, if it
is described by a Vielbein instead of by a metric, has a gauge
freedom, as it transforms under local Lorentz transformations. In
theories with fermions, Viebeins are necessary to work with the
spinorial fields in curved space time.

This problem was first noticed and solved by Jacobson and Mohd in
Ref.~\cite{Jacobson:2015uqa} for the Einstein-Hilbert action written
in terms of the Vielbein. The solution consists in going back to the
basic formalism of \cite{Lee:1990nz,Wald:1993nt} and deal carefully
with the gauge (local Lorentz) symmetry. In practice, this means
taking into account the gauge transformations induced by the
diffeomorphisms on the Vielbein. This can be done, for instance, by
defining a Lorentz-covariant Lie derivative (\textit{Lie-Lorentz
  derivative}) which can be decomposed into a standard Lie derivative
and a local Lorentz transformation and which, apart from being
covariant under Local lorentz transformations, vanishes identically
when the diffeomorphism is an isometry of the metric (see
Refs.~\cite{Ortin:2002qb,Ortin:2015hya}\footnote{See also
  Ref.~\cite{kn:FF} for a more mathematically rigorous point of view.}
which build on earlier work by Lichnerowicz, Kosmann and others
\cite{kn:Lich,kn:Kos,kn:Kos2,Hurley:cf,Vandyck:ei,Vandyck:gc,Figueroa-O'Farrill:1999va}).
The Lie-Lorentz derivative has been recently used to extend the proof
of the first law of black mechanics to supergravity, including the
spinorial fields, in Ref.~\cite{Aneesh:2020fcr}.

A more mathematically rigorous (and complicated) treatment based on
the theory of principal bundles, that also applied to Yang-Mills
fields, was given by Prabhu in
Ref.~\cite{Prabhu:2015vua}.\footnote{See also
  Ref.~\cite{Hajian:2015xlp} for a different point of view on this
  problem.}  Apart from the mathematical complexity, this approach
cannot be used to handle higher-rank form fields such as the
Kalb-Ramond (KR) field. For this reason, in Ref.~\cite{Elgood:2020svt}
we proposed a simpler alternative, based on the construction of
covariant Lie derivatives of all the fields with gauge freedom (a
Maxwell field in the case of Ref.~\cite{Elgood:2020svt}). This
construction is based on the introduction of \textit{momentum maps}
\cite{Ortin:2015hya,Bandos:2016smv} which play a crucial role in this
paper and which we will define later. The Lie-Lorentz derivative can
also be seen as based on the definition of a Lorentz momentum
map.\footnote{In Refs.~\cite{Frodden:2017qwh,Frodden:2019ylc},
  momentum maps emerge as ``improved gauge transformations''.}

In Ref.~\cite{Elgood:2020mdx} we have shown how to use momentum maps
to construct covariant Lie derivatives in the Heterotic Superstring
Effective action compactified in a torus at zeroth order in
$\alpha'$. The KR field of that theory contains Abelian Chern-Simons
terms\footnote{Only the Kaluza-Klein and winding vector fields appear
  there at zeroth order in $\alpha'$.} which induce Nicolai-Townsend
transformations of the 2-form \cite{Nicolai:1980td}. These terms
modify the definitions of the conserved charges which ultimately
appear in the first law of black hole mechanics along the lines of the
classical
Refs.~\cite{Regge:1974zd,Abbott:1981ff,Barnich:2001jy,Barnich:2003xg}.

In this paper we are going to use the same technique quite extensively
to deal with the variety of fields and couplings that occur in the
Heterotic Superstring effective action at first order in $\alpha'$ and
prove the first law of black hole mechanics, identifying the
entropy. As we are going to see, the entropy formula obtained is
manifestly gauge-invariant and contains only terms which are known and
can be computed explicitly. This is the first entropy formula proposed
for this theory that satisfies all this properties. It allows us to
compute reliably the entropy of black hole solutions to first order in
$\alpha'$ and compare the result with the entropy computed through
microstate counting. As we will show in the last section, it gives the
same results as the non-gauge-invariant formulae used in
Refs.~\cite{Cano:2019ycn,Elgood:2020xwu,Ortin:2020xdm} in certain
basis.\footnote{These results differ slightly from the results
  obtained in Refs.~\cite{Cano:2018qev,Cano:2018brq} using the
  Iyer-Wald prescription in the higher-dimensional action before
  dimensional reduction. As pointed out in Ref.~\cite{Faedo:2019xii},
  the dependence on the Riemann tensor changes after dimensional
  reduction and the formulae in
  Refs.~\cite{Cano:2019ycn,Elgood:2020xwu,Ortin:2020xdm} have been
  found using the dimensionally-reduced action. The formula that we
  give here does not suffer of any of these problems. See the
  discussion in Section~\ref{sec-discussion}.} This confirms the
values of the entropies obtained in those references, and shows why,
in spite of the manifest deficiencies of the entropy formulae used, we
obtained the right result.

A very interesting aspect of the momentum maps is that they are
related to the zeroth law of black hole mechanics and its
generalizations.\footnote{This was first noticed by Prabhu, albeit in
  a completely different language \cite{Prabhu:2015vua}.} In the
simplest case, the momentum map associated to a Maxwell field can be
interpreted as the electrostatic potential.\footnote{The Maxwell
  momentum map is defined in a gauge invariant form, and so is the
  electostatic potential. This is in contrast wit the standard
  definitions of the electrostatic potential used in the literature.}
The \textit{generalized zeroth law} states that it is constant over
the black hole horizon \cite{Bardeen:1973gs}. The horizon's surface
gravity, which is the subject of the zeroth law, is also related to
the Lorentz momentum map. For higher-rank fields, Copsey and Horowitz
\cite{Copsey:2005se} and, afterwards, Comp\`ere \cite{Compere:2007vx}
proved a restricted form of the generalized zeroth law (restricted
because it refers only to the bifurcation sphere) which follows from
the closedness of certain differential form on it. In
Ref.~\cite{Elgood:2020mdx} we proved that these closed forms are
related to the momentum maps and we will call these statements
\textit{restricted generalized zeroth laws}. Here we will extend the
results of Ref.~\cite{Elgood:2020mdx} to YM and KR fields and to the more
complicated couplings of the Heterotic Superstring effective action at
first order in $\alpha'$.\footnote{Some of these couplings have been
  discussed before in the literature, specially in
  Ref.~\cite{Tachikawa:2006sz} (see also references therein). See the
  discussion in Section~\ref{sec-discussion}.}

The restricted generalized zeroth laws play a crucial role in the
proof of the first law and in the identification of the entropy and
they are intimately related to the definitions of conserved
charges. In Wald's formalism, the entropy is identified only after the
terms $\sim\Phi \delta \mathcal{Q}$ have been identified in the first
law. As in Ref.~\cite{Elgood:2020mdx}, this identification requires
the addition and subtraction of several terms as demanded by the
definitions of the charges $\mathcal{Q}$ and the potentials $\Phi$ on
account of the restricted generalized zeroth laws. However, in this
case, some of the terms added and subtracted will be shown to
contribute to the entropy.
 
This paper is organized as follows: in
Section~\ref{sec-firstorderaction} we introduce the effective action
of the Heterotic Superstring to first order in $\alpha'$ and find how
it changes under an arbitrary variation of the fields, which allows us
to determine the equations of motion. In
Section~\ref{sec-firstordertrans} we study how the fields change under
gauge and general coordinate transformations. We construct variations
of the fields that vanish when the parameters of the transformations
generate a symmetry of the field configuration and we find the
integrals that give the associated conserved charges. The conserved
charge associated to the invariance under diffeomorphisms is the
Wald-Noether charge. As we have discussed, the correct identification
of the conserved charges is essential to obtain for the correct
identification of the entropy in the first law.  In
Section~\ref{sec-zerothlaws} we discuss the restricted generalized
zeroth laws of this theory, which also play an essential role in the
proof of the first law.  In Section~\ref{sec-firstlaw} we prove the
first law using the results obtained in the previous sections, which
leads us to identify the Wald entropy formula in
Section~\ref{sec-entropy}. Section~\ref{sec-discussion} contains a
discussion of our results, comparing them with the existing
literature.

\section{The HST effective action  at first order in $\alpha'$}
\label{sec-firstorderaction}

The Heterotic Superstring effective action can be described at first
order in $\alpha'$ as follows \cite{Bergshoeff:1989de}:\footnote{We
  use the conventions of Ref.~\cite{Ortin:2015hya}, reviewed for the
  zeroth-order case in Ref.~\cite{Elgood:2020mdx}. In particular, the
  relation with the fields in Ref.~\cite{Bergshoeff:1989de} can be
  found in Ref.~\cite{Fontanella:2019avn}.}  we start by defining the
zeroth-order KR field strength $H^{(0)}$ and its components
$H^{(0)}{}_{\mu\nu\rho}$ as

\begin{equation}
  \label{eq:H0def}
  H^{(0)} \equiv dB
  =
  \tfrac{1}{3!}H_{\mu\nu\rho}dx^{\mu}\wedge dx^{\mu}\wedge dx^{\rho}\, ,
\end{equation}

\noindent
where $B=\tfrac{1}{2}B_{\mu\nu}dx^{\mu}\wedge dx^{\mu}$ is the KR
2-form potential. Then, if $\omega^{ab}=\omega_{\mu}{}^{ab}dx^{\mu}$
is the Levi-Civita spin connection,\footnote{If
  $e^{a}= e^{a}{}_{\mu}dx^{\mu}$ are the Vielbein, the spin connection
  is defined to satisfy the Cartan structure equation
  $\mathcal{D}e^{a}\equiv de^{a}-\omega^{a}{}_{b}\wedge e^{b}=0$.} we
define the zeroth-order torsionful spin connections\footnote{We denote
  by $\imath_{a} A$ the inner product of
  $e_{a}\equiv e_{a}{}^{\mu}\partial_{\mu}$
  ($e_{a}{}^{\mu}e^{b}{}_{\mu} = \delta^{a}{}_{b}$) with the
  differential form $A$. If $A$ is a $p$-form with components
  $A_{\mu_{1}\cdots\mu_{p}}$, $\imath_{a}A$ is the $(p-1)$ form with
  components $e_{a}{}^{\nu}A_{\nu\mu_{1}\cdots\mu_{p-1}}$.}

\begin{equation}
  \label{eq:Omegadef}
\Omega^{(0)}_{(\pm)\, ab}
=
\omega_{ab}
\pm
\tfrac{1}{2}\imath_{b}\imath_{a}H^{(0)}\, ,
\end{equation}

\noindent
and their corresponding zeroth-order curvature 2-forms and
Chern-Simons 3-forms

\begin{subequations}
  \label{eq:R0def}
  \begin{align}
R^{(0)}_{(\pm)}{}^{ab}
 & \equiv 
            d\Omega^{(0)}_{(\pm)}{}^{ab}
            -\Omega^{(0)}_{(\pm)}{}^{a}{}_{c}\wedge \Omega^{(0)}_{(\pm)}{}^{cb}\, ,  
\\
  & & \nonumber \\
  \label{eq:oL0def}
\omega^{(0)}_{(\pm)}
  & = 
        R^{(0)}_{(\pm)}{}^{a}{}_{b}\wedge \Omega^{(0)}_{(\pm)}{}^{b}{}_{a}
        +\tfrac{1}{3} \Omega^{(0)}_{(\pm)}{}^{a}{}_{b}\wedge
        \Omega^{(0)}_{(\pm)}{}^{b}{}_{c} \wedge
        \Omega^{(0)}_{(\pm)}{}^{c}{}_{a}\, .
  \end{align}
\end{subequations}

Next, we  define the gauge field strength 2-form and the Chern-Simons 3-forms
for the YM field $A^{A}=A^{A}{}_{\mu}dx^{\mu}$ by

\begin{eqnarray}
  \label{eq:FAmn}
F^{A}
& = & 
dA^{A} +\tfrac{1}{2}f_{BC}{}^{A}A^{B}\wedge A^{C}\, , 
\\
& & \nonumber \\
\omega^{\rm YM}
  & = &
        F_{A}\wedge A^{A} -\tfrac{1}{6}f_{ABC}A^{A}\wedge A^{B} \wedge A^{C}\, ,
\end{eqnarray}

\noindent
where we have lowered the adjoint group indices $A,B,C,\ldots$ in the
structure constants $f_{AB}{}^{C}$ and gauge fields using the
Killing metric.

Then, we can define the first-order KR field strength 3-form as

\begin{equation}
  \label{eq:H1def}
H^{(1)}
\equiv 
H^{(0)}
+\frac{\alpha'}{4}\left(\omega^{\rm YM}+\omega^{(0)}_{(-)}\right)\, .  
\end{equation}

Its Bianchi identity takes the well-known form

\begin{equation}
  \label{eq:BianchiH1}
  dH^{(1)}
  =
  \frac{\alpha'}{4}\left(F_{A}\wedge F^{A}
    +R^{(0)}_{(-)}{}^{a}{}_{b}\wedge R^{(0)}_{(-)}{}^{b}{}_{a} \right)\, .
\end{equation}

Having made these definitions and adding the dilaton field $\phi$, we
can write the Heterotic Superstring effective action to first-order in
$\alpha'$ as

\begin{equation}
\label{eq:heterotic1diff}
\begin{aligned}
  S^{(1)}[e^{a},B,A^{A},\phi]
  & =
  \frac{g_{s}^{(d)\, 2}}{16\pi G_{N}^{(d)}} \int e^{-2\phi}
  \left[ (-1)^{d-1} \star (e^{a}\wedge e^{b}) \wedge R_{ab}
    -4d\phi\wedge \star d\phi
  \right.
  \\
  & \\
  &
  \left.
    +\tfrac{1}{2}H^{(1)}\wedge \star H^{(1)}
    +(-1)^{d}\frac{\alpha'}{4}\left(F_{A}\wedge \star F^{A}
      +R^{(0)}_{(-)}{}^{a}{}_{b}\wedge \star R^{(0)}_{(-)}{}^{b}{}_{a}\right)  
  \right]
  \\
  & \\
  & \equiv
  \int  \mathbf{L}^{(1)}\, .
\end{aligned}
\end{equation}

Although this action is defined in 10 dimensions, we have left the
dimension arbitrary $(d)$ because that allows us to use the results in
other dimensions after trivial dimensional reduction on a torus. 
In this action, $G_{N}{}^{(d)}$ is the $d$-dimensional Newton constant
and $g_{s}^{(d)}$ is the $d$-dimensional string coupling constant,
identified with the vacuum expectation value of the exponential of the
$d$-dimensional dilaton field $g_{s}^{(d)}=<e^{\phi}>$. In solutions
such as black holes that asymptote to a vacuum solution at infinity
$e^{\phi}\rightarrow e^{\phi_{\infty}}=<e^{\phi}>=g_{s}^{(d)}$.

This is a very complex action.  Due to this complexity and to the
lemma proven in Ref.~\cite{Bergshoeff:1989de} which we will explain
later, it is convenient to perform a general variation of the action
in two steps: first, we only vary the action with respect to the
\textit{explicit} occurrences of the fields, where we define
``explicit occurrences'' as those which do not take place in the
torsionful spin connection $\Omega^{(0)}_{(-)}$. Then, we vary the
action with respect to the occurrences of the fields via
$\Omega^{(0)}_{(-)}$ using the chain rule. All the occurrences of the
dilaton and YM fields are explicit, but those of the Vielbein and KR
field are not, because they (and only they) are present in
$\Omega^{(0)}_{(-)}$.

Thus, setting $g_{s}^{(d)\, 2}(16\pi G_{N}^{(d)})^{-1}=1$ for the time
being in order to simplify the formulae, we find that under a general
variation of the ``explicit'' occurrences of the fields, the action
transforms as follows:

\begin{equation}
  \label{eq:deltaS1exp}
  \begin{aligned}
    \delta_{\rm exp} S^{(1)}
    & =
    \int \left\{ \mathbf{E}^{(1)}_{{\rm exp}\,
        a}\wedge \delta e^{a} +\mathbf{E}^{(1)}_{{\rm exp}\, B}\wedge \delta B
      +\mathbf{E}^{(1)}_{\phi}\delta\phi +\mathbf{E}^{(1)}_{A}\delta A^{A}
    \right.
    \\
    & \\
    &
    \hspace{.5cm}
    \left.
      +d\mathbf{\Theta}^{(1)}_{\rm exp}(\varphi,\delta\varphi) \right\}\, ,
\end{aligned}
\end{equation}

\noindent
where $\varphi$ stands for all the fields of the theory,

\begin{subequations}
  \begin{align}
    \label{eq:Eaexp}
  \mathbf{E}^{(1)}_{{\rm exp}\, a}
  & =
    e^{-2\phi} \imath_{a}\star (e^{c}\wedge e^{d})\wedge R_{cd}
    -2\mathcal{D}(\imath_{b}de^{-2\phi})\wedge \star(e^{b}\wedge e^{c})g_{ca}
    \nonumber \\
    & \nonumber \\
    &
      \hspace{.5cm}
      +(-1)^{d-1}4e^{-2\phi}
      \left(\imath_{a}d\phi \star d\phi+d\phi\wedge \imath_{a}\star d\phi\right)
     \nonumber \\
    & \nonumber \\
    &
      \hspace{.5cm}
      +\frac{(-1)^{d}}{2}e^{-2\phi}
      \left(\imath_{a}H^{(1)}\wedge \star H^{(1)}+H^{(1)}\wedge \imath_{a}\star H^{(1)}\right)
    \nonumber \\
    & \nonumber \\
    &
       \hspace{.5cm}
   +\frac{\alpha'}{4}e^{-2\phi}\left(\imath_{a}F_{A}\wedge \star F^{A}-F_{A}\wedge
      \imath_{a}\star F^{A}
      \right.
      \nonumber \\
    & \nonumber \\
    &
       \hspace{.5cm}
      \left.
      +\imath_{a}R^{(0)}_{(-)}{}^{b}{}_{c}\wedge \star R^{(0)}_{(-)}{}^{c}{}_{b}
      -R^{(0)}_{(-)}{}^{b}{}_{c}\wedge \imath_{a}\star R^{(0)}_{(-)}{}^{c}{}_{b}\right)
      \\
    & \nonumber \\
    \label{eq:EBexp}
    \mathbf{E}^{(1)}_{{\rm exp}\, B}
    & =
      -d\left(e^{-2\phi}\star H^{(1)}\right)\, ,
          \\
    & \nonumber \\
    \label{eq:Ephi1}
    \mathbf{E}^{(1)}_{\phi}
    & =
      8d\left(e^{-2\phi}\star d\phi\right)-2\mathbf{L}^{(1)}\, ,
          \\
    & \nonumber \\
    \label{eq:EA}
    \mathbf{E}^{(1)}_{A}
    & =
      -\frac{\alpha'}{2}
      \left\{\mathcal{D}\left(e^{-2\phi} \star F_{A} \right)
      +(-1)^{d}e^{-2\phi} \star H^{(0)}\wedge F_{A}\right\}
      -\frac{\alpha'}{4} \mathbf{E}^{(1)}_{{\rm exp}\, B}\wedge A_{A}\, ,
 \end{align}
 \end{subequations}
 
 \noindent
 and

 \begin{equation}
   \label{eq:Thetaexp}
  \begin{aligned}
    \mathbf{\Theta}^{(1)}_{\rm exp}(\varphi,\delta\varphi)
    & =
    -e^{-2\phi}\star (e^{a}\wedge e^{b})\wedge \delta \omega_{ab}
    +2\imath_{a}de^{-2\phi}\star(e^{a}\wedge e^{b})\wedge \delta e_{b}
    -8e^{-2\phi}\star d\phi\delta\phi
    \\
    & \\
    &
    \hspace{.5cm}
    +e^{-2\phi}\star H^{(1)}\wedge \delta B
+\frac{\alpha'}{2}e^{-2\phi}\left(\star F_{A} -\tfrac{1}{2}\star H^{(1)}\wedge
  A_{A}\right)\wedge \delta A^{A}\, .
  \end{aligned}
\end{equation}

\noindent
An alternative form of the YM equations that arises in the calculations is

\begin{equation}
\label{eq:EA2}
\mathbf{E}^{(1)}_{A}
 =
-\frac{\alpha'}{2}
\mathcal{D}\left(e^{-2\phi} \star F_{A} 
-e^{-2\phi} \star H^{(0)}\wedge A_{A}\right)
+(-1)^{d-1}\frac{\alpha'}{4}e^{-2\phi} \star H^{(0)} \wedge dA_{A}\, .
\end{equation}

Observe that neither the YM equations of motion transform covariantly
nor $\mathbf{\Theta}^{(1)}_{\rm exp}$ is invariant under YM gauge
transformations. For the YM equations this is not a big problem since
the troublesome term is proportional to the KR equation of motion, but
there is no obvious fix for the pre-symplectic potential. Nevertheless,
we will see that, in the end, we will get  gauge-invariant charges
and, in particular a gauge-invariant Wald-Noether charge.

An important property of the HST effective action is that the YM fields
and the torsionful spin connection occur in it exactly on the same footing
\cite{Bergshoeff:1988nn}. The variation of the action with respect to the
torsionful spin connection takes exactly the same form as the YM equation, the
only difference being the group indices and their contractions. Thus,

\begin{equation}
  \label{eq:deltaS1}
  \begin{aligned}
    \delta S^{(1)} &
    = \int \left\{
      \mathbf{E}^{(1)}_{{\rm exp}\, a}\wedge \delta e^{a}
      +\mathbf{E}^{(1)}_{{\rm exp}\, B}\wedge \delta B
      +\mathbf{E}^{(1)}_{\phi}\delta\phi
      +\mathbf{E}^{(1)}_{A} \wedge \delta A^{A}
      +\mathbf{E}^{(1)\, b}{}_{a}\wedge \delta \Omega^{(0)}_{(-)}{}^{a}{}_{b}
    \right.
    \\
    & \\
    &
    \hspace{.5cm}
    \left.
      +d\mathbf{\Theta}^{(1)}(\varphi,\delta\varphi) \right\}\, ,
\end{aligned}
\end{equation}

\noindent
where the variation with respect to the torsionful spin connection is
given by

\begin{equation}
    \label{eq:Eab}
    \mathbf{E}^{(1)\, b}{}_{a}
 =
      -\frac{\alpha'}{2}
      \left\{\mathcal{D}_{(-)}\left(e^{-2\phi} \star R^{(0)}_{(-)}{}^{b}{}_{a} \right)
      +(-1)^{d}e^{-2\phi} \star H^{(0)}\wedge R^{(0)}_{(-)}{}^{b}{}_{a}\right\}
      -\frac{\alpha'}{4} \mathbf{E}^{(1)}_{{\rm exp}\, B}\wedge \Omega^{(0)}_{(-)}{}^{b}{}_{a}\, ,
\end{equation}

\noindent
or

\begin{equation}
\label{eq:Eab2}
    \mathbf{E}^{(1)\, b}{}_{a}
 =
      -\frac{\alpha'}{2}
      \mathcal{D}_{(-)}\left(e^{-2\phi} \star R^{(0)}_{(-)}{}^{b}{}_{a} 
      -e^{-2\phi} \star H^{(0)}\wedge \Omega^{(0)\, b}_{(-)}{}_{a}\right)
      +(-1)^{d-1}\frac{\alpha'}{4}\star H^{(0)}\wedge d\Omega^{(0)\, b}_{(-)}{}_{a}\, ,
\end{equation}

\noindent
and the pre-symplectic $(d-1)$-form is given by

\begin{equation}
  \label{eq:Theta1}
\begin{aligned}
    \mathbf{\Theta}^{(1)}(\varphi,\delta\varphi)
    & =
        \mathbf{\Theta}^{(1)}_{\rm exp}(\varphi,\delta\varphi)
    +\frac{\alpha'}{2}e^{-2\phi}\left(\star R^{(0)}_{(-)}{}^{b}{}_{a}
      -\tfrac{1}{2}\star H^{(1)}\wedge \Omega^{(0)}_{(-)}{}^{b}{}_{a}\right)
    \wedge \delta  \Omega^{(0)}_{(-)}{}^{a}{}_{b}\, ,
  \end{aligned}
\end{equation}

\noindent
with $\mathbf{\Theta}^{(1)}_{\rm exp}(\varphi,\delta\varphi)$ 
given in Eq.~(\ref{eq:Thetaexp}).

The parallelism between the YM and torsionful spin connection terms
also leads to the same problems of non-covariance of
$\mathbf{E}^{(1)\, b}{}_{a}$ and non-invariance of the additional term
in $\mathbf{\Theta}^{(1)}$.

An important difference between the equations of motion of these two
connections is that, according to the lemma proven in
Ref.~\cite{Bergshoeff:1989de}, $\mathbf{E}^{(1)\, a}{}_{b}$ is
proportional to $\alpha'$ and to a combination of the zeroth-order
equations $\mathbf{E}^{(0)}_{a}, \mathbf{E}^{(0)}_{B}$ and
$\mathbf{E}^{(0)}_{\phi}$.  This means that field configurations that
solve the equations $\mathbf{E}^{(1)}_{{\rm exp}\, a}=0$,
$\mathbf{E}^{(1)}_{{\rm exp}\, B}=0$, $\mathbf{E}^{(1)}_{\phi}=0$ and
$\mathbf{E}^{(1)}_{A}=0$ are solutions of the complete first-order
equations, to that order in $\alpha'$. This crucial property
effectively reduces the degree of the differential equations to $2$,
avoiding the problems that arise with dynamical equations that involve
derivatives of the fields of higher order.

\section{Variations of the fields}
\label{sec-firstordertrans}

It is convenient to start by describing the gauge transformations of
the fields and the associated Noether identities to be able to compute
the associated conserved charges. Afterwards, we will discuss the
transformations of the fields under diffeomorphisms and the associated
Wald-Noether charge.

\subsection{Gauge transformations}
\label{sec-gauge}

The fields occurring in the effective action
Eq.~(\ref{eq:heterotic1diff}) transform under 3 kinds of gauge
transformations:

\begin{enumerate}

\item KR gauge transformations with 1-form parameter $\Lambda$,
  $\delta_{\Lambda}$, which only act on $B$.
\item YM gauge transformations with parameter $\chi^{A}$,
  $\delta_{\chi}$, which act on the YM fields and on $B$ as
  Nicolai-Townsend transformations.
\item Local Lorentz transformations with parameter $\sigma^{ab}$,
  $\delta_{\sigma}$, which act on the Vielbein and induce
  transformations of spin connections and curvature and which also act
  on $B$ as Nicolai-Townsend transformations.

\end{enumerate}

The transformation rules are

\begin{subequations}
  \begin{align}
    \delta_{\sigma} e^{a}
    & =
    \sigma^{a}{}_{b}e^{b}\, ,
    \\
    & \nonumber \\
     \delta_{\chi} A^{A}
    & =
    \mathcal{D}\chi^{A}
    \equiv
    d\chi^{A}+f_{BC}{}^{A}A^{B}\chi^{C}\, ,
    \\
    & \nonumber \\
    \delta B
    & =
    (\delta_{\Lambda}+\delta_{\chi}+\delta_{\sigma}) B
      =
    d\Lambda
    -\frac{\alpha'}{4}\chi_{A}dA^{A}
      -\frac{\alpha'}{4}\sigma^{a}{}_{b}d\Omega^{(0)}_{(-)}{}^{b}{}_{a}\, .
  \end{align}
\end{subequations}

The induced local Lorentz transformations of the connections are

\begin{subequations}
  \begin{align}
\delta_{\sigma} \omega^{ab}
    & =
      \mathcal{D}\sigma^{ab}
      =
       d\sigma^{ab}-2\omega^{[a|}{}_{c}\sigma^{c|b]}\, ,
    \\
    & \nonumber \\
\delta_{\sigma} \Omega^{(0)}_{(-)}{}^{ab}
    & =
      \mathcal{D}^{(0)}_{(-)}\sigma^{ab}
      =
      d\sigma^{ab}
      -2\Omega^{(0)}_{(-)}{}^{[a|}{}_{c}\sigma^{c|b]}\, ,
  \end{align}
\end{subequations}

\noindent
and the transformations of the curvatures are

\begin{subequations}
 \begin{align}
      \delta_{\chi} F^{A}
   & =
     -\chi^{B}f_{BC}{}^{A}F^{C}
    \\
    & \nonumber \\
      \delta_{\sigma} R^{ab}
      & =
        2\sigma^{[a|}{}_{c}R^{c|b]}\, .
    \\
    & \nonumber \\
      \delta_{\sigma} R^{(0)}_{(-)}{}^{ab}
      & =
        2\sigma^{[a|}{}_{c}R^{(0)}_{(-)}{}^{c|b]}\, .
  \end{align}
\end{subequations}

Finally, for the sake of completeness and their later use, we quote
the gauge transformations of the Chern-Simons 3-forms

\begin{subequations}
  \begin{align}
    \delta_{\chi} \omega^{\rm YM}
    & =
  \frac{\alpha'}{4}d\left(\chi_{A}dA^{A}\right)\, ,
      \\
    & \nonumber \\
    \delta_{\sigma} \omega^{(0)}_{(-)}
    & =
      +\frac{\alpha'}{4}d\left(\sigma^{a}{}_{b}d\Omega^{(0)}_{(-)}{}^{b}{}_{a}\right)\, ,
  \end{align}
\end{subequations}

\noindent
and the Ricci identities

\begin{subequations}
  \begin{align}
  \label{eq:YMRicci}
  \mathcal{D}\mathcal{D}\chi^{A}
  & =
    -f_{BC}{}^{A}\chi^{B}F^{C}
    =
    \delta_{\chi}F^{A}\, ,
    \\
    & \nonumber \\
  \mathcal{D}^{(0)}_{(-)}\mathcal{D}^{(0)}_{(-)}\sigma^{ab}
  & =
    -2R^{(0)}_{(-)}{}^{[a|}{}_{c}\sigma^{c|b]}
    =
    \delta_{\sigma}R^{(0)}_{(-)}{}^{ab}\, .
  \end{align}
\end{subequations}

The exact invariance of the action $S^{(1)}$ in
Eq.~(\ref{eq:heterotic1diff}) under the above gauge transformations
leads, in a rather trivial way, to the following Noether identities
\cite{Fontanella:2019avn}

\begin{subequations}
  \begin{align}
  \label{eq:noetheridentity1}
    d\mathbf{E}^{(1)}_{{\rm exp}\, B}
    & = 0\, ,
    \\
    & \nonumber \\
  \label{eq:noetheridentity2}
    \mathcal{D}  \mathbf{E}^{(1)}_{A}
    +(-1)^{d-1}\frac{\alpha'}{4}\mathbf{E}^{(1)}_{{\rm exp}\, B}\wedge dA_{A}
  & =
  0\, ,
    \\
    & \nonumber\\
    \label{eq:noetheridentity3}
\mathcal{D}^{(0)}_{(-)}\mathbf{E}^{(1)}{}_{b}{}^{a}
    +(-1)^{d-1}\frac{\alpha'}{4}\mathbf{E}^{(1)}_{{\rm exp}\, B}\wedge
    d\Omega^{(0)}_{(-)\, b}{}^{a}
&  =
  0\, ,
    \\
    & \nonumber\\
    \label{eq:noetheridentity4}
    \mathbf{E}_{\rm exp}^{(1)\, [a}\wedge e^{b]}
    +\frac{\alpha'}{4}
    \mathbf{E}^{(1)}_{{\rm exp}\, B}\wedge d \Omega^{(0)\, ab}
    +(-1)^{d-1}\mathcal{D}^{(0)}_{(-)}\mathbf{E}^{(1)\, ab}
    & =
      0\, .
\end{align}
\end{subequations}

Eq.~(\ref{eq:noetheridentity3}) is just a particular case of
Eq.~(\ref{eq:noetheridentity2}) with adjoint Lorentz
indices. Furthermore, the last two identities imply the symmetry of
the Einstein equation, which in the language f differential forms and
Vielbeins, is expressed in the form

\begin{equation}
    \label{eq:noetheridentity5}
    \mathbf{E}_{\rm exp}^{(1)\, [a}\wedge e^{b]}
    =
    0\, .
\end{equation}

\subsection{Gauge charges}
\label{sec-gaugecharges}

For ths sake of simplicity, we are going to start by the charge
associated to the $\delta_{\Lambda}$ transformations, that we are
going to call Kalb-Ramond charge.

\subsubsection{Kalb-Ramond charge}
\label{sec-KRcharge}

Let us consider the transformation of the action
Eq.~(\ref{eq:heterotic1diff}) under the gauge transformations
$\delta_{\Lambda}$. Taking into account that this symmetry only acts
on $B$,\footnote{We consider the variation of the torsionful spin
  connection to be zero under this transformation.}
Eqs.~(\ref{eq:deltaS1}) and (\ref{eq:Theta1}) we get

\begin{equation}
  \label{eq:deltaLambdaS1}
  \begin{aligned}
    \delta_{\Lambda} S^{(1)}
    & =
    \int \left\{
      \mathbf{E}^{(1)}_{{\rm exp}\, B}\wedge d\Lambda
      +d\left[
        e^{-2\phi}\star H^{(1)}\wedge  d\Lambda
\right]
\right\}\, .
\end{aligned}
\end{equation}

\noindent
Integrating by parts the first term and using the Noether identity
Eq.~(\ref{eq:noetheridentity1})

\begin{equation}
  \label{eq:deltaLambdaS1-2}
  \begin{aligned}
    \delta_{\Lambda} S^{(1)}
    & =
    \int d\left\{
        (-1)^{d}\mathbf{E}^{(1)}_{{\rm exp}\, B}\wedge \Lambda
        +e^{-2\phi}\star H^{(1)}\wedge  d\Lambda\right\}
      \equiv
      \int d\mathbf{J}[\Lambda]\, .
\end{aligned}
\end{equation}

Since $ \delta_{\Lambda} S^{(1)}=0$, the integrand must vanish, which
means that $\mathbf{J}[\Lambda]$ must be locally exact. Indeed,

\begin{equation}
  \mathbf{J}[\Lambda] = d\mathbf{Q}[\Lambda]\, ,
  \,\,\,\,\,
  \text{with}
  \,\,\,\,\,
  \mathbf{Q}[\Lambda]
  = \Lambda\wedge\left(e^{-2\phi}\star H^{(1)}\right)\, .
\end{equation}

\noindent
Integrating the $(d-2)$-form $\mathbf{Q}[\Lambda]$ over
$(d-2)$-dimensional compact surfaces $\mathcal{S}_{d-2}$ for
$\Lambda$s that leave invariant the KR field $B$ we get conserved
charges associated to those $\Lambda$s.  These $\Lambda$s are simply
closed 1-forms.\footnote{Here we follow
  Refs.~\cite{Copsey:2005se,Compere:2007vx}. This discussion is
  identical to the discussion we made for the zeroth-order case in
  Ref.~\cite{Elgood:2020mdx}.} The Hodge decomposition theorem allows
us to write each of them as the sum of an exact and a harmonic form
that we denote by $\Lambda_{e}$ and $\Lambda_{h}$, respectively.
On-shell, the exact form $\Lambda_{e}=d\lambda$ will not contribute to
the integral and the charge will be given by

\begin{equation}
  \mathcal{Q}(\Lambda_{h}) = \int_{\mathcal{S}_{d-2}}
  \Lambda_{h}\wedge \left(e^{-2\phi}\star H \right)\, .
\end{equation}

\noindent
Now we can use duality between homology and cohomology: if
$C_{\Lambda_{h}}$ is the $(d-3)$-cycle dual to $\Lambda_{h}$ we arrive
at the charges

\begin{equation}
  \label{eq:QLambdacharge}
  \mathcal{Q}(\Lambda_{h})
  =
  -\frac{g^{(d)\, 2}_{s}}{16\pi G_{N}^{(d)}}\int_{C_{\Lambda_{h}}}e^{-2\phi}\star H\, ,
\end{equation}

\noindent
where we have recovered the factor of
$g^{(d)\, 2}_{s}(16\pi G_{N}^{(d)})^{-1}$ and added a conventional
sign.

\subsubsection{Yang-Mills charge}
\label{sec-YMcharge}

Now, let us consider the charges associated to the YM gauge
transformations $\delta_{\chi}$. Again, from Eqs.~(\ref{eq:deltaS1})
and (\ref{eq:Theta1}), taking into account that this symmetry acts on
the YM fields $A^{A}$ but also on the KR 2-form $B$, we have

\begin{equation}
  \label{eq:deltachiS1}
  \begin{aligned}
    \delta_{\chi} S^{(1)} &
    = \int \left\{
      \mathbf{E}^{(1)}_{{\rm exp}\, B}\wedge \delta_{\chi} B
      +\mathbf{E}^{(1)}_{A} \wedge \delta_{\chi} A^{A}
    \right.
    \\
    & \\
    &
    \hspace{.5cm}
    \left.
      +d\left[e^{-2\phi}\star H^{(1)}\wedge \delta_{\chi} B
+\frac{\alpha'}{2}e^{-2\phi}\left(\star F_{A} -\tfrac{1}{2}\star H^{(1)}\wedge
  A_{A}\right)\wedge \delta_{\chi} A^{A}
      \right]\right\}\, .
\end{aligned}
\end{equation}

The parameters $\chi^{A}$ that we will use are those that preserve the
field configuration, leaving $A^{A}$ and $B$ invariant. The YM fields
are left invariant by covariantly constant $\chi^{A}$s,
\textit{i.e.}~$\chi^{A}$s that we will denote by $\kappa^{A}$
satisfying

\begin{equation}
\mathcal{D}\kappa^{A}=0\, .
\end{equation}

\noindent
We can call these parameters \textit{vertical Killing vector fields}
from he principal bundle point of view, with the standard Killing
vectors of the base manifold playing the r\^ole of \textit{horizontal
  Killing vector fields}.

The integrability condition of the vertical Killing vector equation
is, according to Eq.~(\ref{eq:YMRicci}),

\begin{equation}
  \delta_{\kappa}F^{A}
  =
  -f_{BC}{}^{A}\kappa^{B}F^{C}
  =
  0\, ,
\end{equation}

\noindent
so they also leave invariant the field strengths, as expected.

The vertical Killing vector fields $\kappa^{A}$s will not leave $B$
invariant, though, but we can rewrite the transformation in the form

\begin{equation}
  \delta_{\chi}B
 =
  -\frac{\alpha'}{4}\kappa_{A}dA^{A}
  =
-\frac{\alpha'}{2}\kappa_{A}F^{A}
  +d\left(\frac{\alpha'}{4}\kappa_{A}A^{A}\right)\, .
\end{equation}

\noindent
Now we observe that, due to the YM Bianchi identity $\mathcal{D}F^{A}=0$,
$\kappa_{A}F^{A}$ is a closed 2-form and, locally, there is a 1-form
$\Psi_{\kappa}$  such that

\begin{equation}
  \label{eq:Psikappa}
d\Psi_{\kappa} = -\kappa_{A}F^{A}\, ,
\end{equation}

\noindent
and which we will call \textit{vertical YM momentum
  map}.\footnote{Compare this equation with the equation satisfied by
  the standard (horizontal) YM momentum map Eq.~(\ref{eq:PA}).}

Then, we define the parameter of a compensating $\Lambda$ transformation

\begin{equation}
  \label{eq:Lambdachidef}
  \Lambda_{\chi}
  =
  -\frac{\alpha'}{2}\Psi_{\chi}
  -\frac{\alpha'}{4}\chi_{A}A^{A}\, ,
\end{equation}

\noindent
where $\Psi_{\chi}$ is a 1-form such that, when $\chi^{A}=\kappa^{A}$
(\textit{i.e.}~when it is a vertical Killing vector field), it
satisfies Eq.~(\ref{eq:Psikappa}). Combining the original
$\delta_{\chi}$ transformation with the compensating
$\delta_{\Lambda_{\chi}}$ transformation we find a new
$\delta_{\chi}B$ that vanishes for covariantly constant $\chi^{A}$s:

\begin{equation}
  \label{eq:deltachiBmodified}
  \delta_{\chi}B
  \equiv
-\frac{\alpha'}{2}\left(d\Psi_{\chi}+\chi_{A}F^{A}\right)
  -\frac{\alpha'}{4}\mathcal{D}\chi_{A}\wedge A^{A}\, .
\end{equation}

The vanishing of $\delta_{\chi}B$ for covariantly constant $\chi^{A}$s is
gauge invariant because

\begin{equation}
\delta_{\chi'}\delta_{\chi} \sim \mathcal{D}\chi\, .
\end{equation}

Substituting the transformation Eq.~(\ref{eq:deltachiBmodified}) and the
standard gauge transformation of the YM fields into Eq.~(\ref{eq:deltachiS1})
we get

\begin{equation}
  \label{eq:deltachiS1-2}
  \begin{aligned}
    \delta_{\chi} S^{(1)} &
    = \int \left\{
      \mathbf{E}^{(1)}_{A} \wedge \mathcal{D}\chi^{A}
      +\mathbf{E}^{(1)}_{{\rm exp}\, B}\wedge
      \left[-d\left(\frac{\alpha'}{2}\Psi_{\chi}
          +\frac{\alpha'}{4}\chi_{A}A^{A} \right) -\frac{\alpha'}{4}\chi_{A}dA^{A}\right]
    \right.
    \\
    & \\
    &
    \hspace{.5cm}
    +d\left\{e^{-2\phi}\star H^{(1)}\wedge
            \left[-d\left(\frac{\alpha'}{2}\Psi_{\chi}
                +\frac{\alpha'}{4}\chi_{A}A^{A} \right)
              -\frac{\alpha'}{4}\chi_{A}dA^{A}\right]
      \right.
    \\
    & \\
    &
    \hspace{.5cm}
    \left.\left.
+\frac{\alpha'}{2}e^{-2\phi}\left(\star F_{A} -\tfrac{1}{2}\star H^{(1)}\wedge
  A_{A}\right)\wedge \mathcal{D}\chi^{A}
      \right\}\right\}\, .
\end{aligned}
\end{equation}

Integrating by parts the first terms and combining the different terms in an
appropriate way we can rewrite the variation in the form 

\begin{equation}
  \label{eq:deltachiS1-3}
  \begin{aligned}
    \delta_{\chi} S^{(1)} &
    = \int \left\{
    (-1)^{d}\chi^{A}\left(\mathcal{D} \mathbf{E}^{(1)}_{A}
      +(-1)^{d-1}\frac{\alpha'}{4}\mathbf{E}^{(0)}_{{\rm exp}\, B}\wedge dA_{A}\right)
    \right.
    \\
    & \\
    &
    \hspace{.5cm}
    -\left(\frac{\alpha'}{2}\Psi_{\chi}
          +\frac{\alpha'}{4}\chi_{A}A^{A} \right)\wedge d\mathbf{E}^{(0)}_{{\rm exp}\, B}
    \\
    & \\
    &
    \hspace{.5cm}
    +d\left\{(-1)^{d-1}\chi^{A}\left(\mathbf{E}^{(1)}_{A}
       +(-1)^{d}\frac{\alpha'}{4}e^{-2\phi}\star H^{(0)}\wedge dA_{A}\right)
      \right.
    \\
    & \\
    &
    \hspace{.5cm}
        -\left(\frac{\alpha'}{2}\Psi_{\chi}
          +\frac{\alpha'}{4}\chi_{A}A^{A}\right)\wedge \mathbf{E}^{(0)}_{{\rm exp}\, B}
    \\
    & \\
    &
    \hspace{.5cm}
      +e^{-2\phi}\star H^{(1)}\wedge
            \left[-d\left(\frac{\alpha'}{2}\Psi_{\chi}
                +\frac{\alpha'}{4}\chi_{A}A^{A} \right)\right]
    \\
    & \\
    &
    \hspace{.5cm}
    \left.\left.
+\frac{\alpha'}{2}e^{-2\phi}\left(\star F_{A} -\tfrac{1}{2}\star H^{(1)}\wedge
  A_{A}\right)\wedge \mathcal{D}\chi^{A}
      \right\}\right\}\, .
\end{aligned}
\end{equation}

\noindent
The terms in the first and second lines vanish identically because of the
Noether identities Eqs.~(\ref{eq:noetheridentity2}) and
(\ref{eq:noetheridentity1}), respectively, and we arrive to

\begin{equation}
  \label{eq:deltachiS1-4}
  \begin{aligned}
    \delta_{\chi} S^{(1)} &
    = \int 
    d\left\{
      (-1)^{d-1}\chi^{A}\left(\mathbf{E}^{(1)}_{A}
       +(-1)^{d}\frac{\alpha'}{4}e^{-2\phi}\star H^{(0)}\wedge dA_{A}\right)
  \right.
    \\
    & \\
    &
    \hspace{.5cm}
        -\left(\frac{\alpha'}{2}\Psi_{\chi}
          +\frac{\alpha'}{4}\chi_{A}A^{A}\right)\wedge \mathbf{E}^{(0)}_{{\rm exp}\, B}
    \\
    & \\
    &
    \hspace{.5cm}
      - d\left(\frac{\alpha'}{2}\Psi_{\chi}
        +\frac{\alpha'}{4}\chi_{A}A^{A} \right)\wedge
      \left(e^{-2\phi}\star H^{(0)}\right)
    \\
    & \\
    &
    \hspace{.5cm}
    \left.
+\frac{\alpha'}{2}e^{-2\phi}\left(\star F_{A} -\tfrac{1}{2}\star H^{(1)}\wedge
  A_{A}\right)\wedge \mathcal{D}\chi^{A}
\right\}
    \\
    & \\
    & \equiv
    \int d\mathbf{J}[\chi]\, .
\end{aligned}
\end{equation}

The same arguments we made in the previous case lead to the existence
of a $(d-2)$-form $\mathbf{Q}[\chi]$ such that
$\mathbf{J}[\chi] = d\mathbf{Q}[\chi]$. The $(d-2)$-form is given by

\begin{equation}
    \mathbf{Q}[\chi]
=
    -(-1)^{d}\frac{\alpha'}{2}\left\{
    e^{-2\phi}\star  \left(-\chi^{A}F_{A}\right)
    +(-1)^{d}\Psi_{\chi}\wedge
      \left(e^{-2\phi}\star H^{(0)}\right)
      \right\}\, .
\end{equation}

For Abelian vector fields the $\kappa^{A}$s are constant and
$\Psi_{\kappa}=\kappa_{A}A^{A}$ (up to a total derivative) and we
recover immediately the $\mathbf{Q}[\chi]$ found in
Ref.~\cite{Elgood:2020mdx}. On the other hand, when we change
$\Psi_{\kappa}$ by a total derivative, $\mathbf{Q}[\kappa]$ is
invariant on-shell up to a total derivative which will not contribute
to the charge which is now given by the integral

\begin{equation}
  \label{eq:YMcharge}
  \mathcal{Q}[\kappa]
  =
  -\frac{g_{s}^{(d)\, 2}}{16 \pi G_{N}^{(d)}}
  \int_{S^{d-2}}
         (-1)^{d}\frac{\alpha'}{2}\left\{e^{-2\phi} \star d\Psi_{\kappa}
      + (-1)^{d}\Psi_{\kappa}\wedge \left(e^{-2\phi} \star H^{(0)}\right)\right\}\, ,
\end{equation}

\noindent
where we have made use of the definition of the vertical momentum map
$\Psi_{\kappa}$ in Eq.~(\ref{eq:Psikappa}).

\subsubsection{Lorentz charge}
\label{sec-Lorentzcharge}
  
Let us now consider local Lorentz transformations. As we have stressed
repeatedly we can treat the local Lorentz transformations and the
torsionful spin connection in parallel to the YM gauge transformations
and the gauge fields. The only difference is the presence of one
additional term in the Lorentz case: the Einstein-Hilbert case. If we
follow the same steps as in the YM case we arrive to

\begin{equation}
  \label{eq:Qsigmad-2}
\mathbf{Q}[\sigma]
=
(-1)^{d-1}e^{-2\phi}\star (e^{a}\wedge e^{b})\sigma_{ab}
-(-1)^{d}\frac{\alpha'}{2}\left\{
    e^{-2\phi}\star  \left(-\sigma^{a}{}_{b}R^{(0)\, b}{}_{a}\right)
    +(-1)^{d}\Pi_{\sigma}\wedge
      \left(e^{-2\phi}\star H^{(0)}\right)
      \right\}\, ,
\end{equation}

\noindent
where $\Pi_{\sigma}$ is a 1-form that becomes a \textit{vertical
  Lorentz momentum map} whan the Lorentz parameter
$\sigma^{a}{}_{b}=\kappa^{a}{}_{b}$, a Lorentz parameter that
generates a symmetry of the field configuration, \textit{i.e.~a
  vertical Killing vector}. This happens when the Vielbein and the
spin connection are left invariant

\begin{subequations}
  \label{eq:kappae=0}
  \begin{align}
    \kappa^{a}{}_{b}e^{b}
    & =
      0\, ,
    \\
    & \nonumber \\
    \label{eq:Dkappa=0}
    \mathcal{D}\kappa^{a}{}_{b}
    & =
      0\, .     
  \end{align}
\end{subequations}

\noindent
These two conditions imply the invariance of the torsion
$\tfrac{1}{2}\imath_{b}\imath_{a}H^{(0)}$ Hence, they also implies the
invariance of the torsionful spin connection
$\Omega^{(0)}_{(-)}{}^{a}{}_{b}$,

\begin{equation}
\mathcal{D}^{(0)}_{(-)}\kappa^{a}{}_{b}=0\, .
\end{equation}

These conditions can be used to modify the transformation of the KR
field so that it is also left invariant, as we did in the YM case. We
just quote the final form:

\begin{equation}
  \label{eq:deltasigmaBmodified}
  \delta_{\sigma}B
  =
  -\frac{\alpha'}{2}\left(d\Pi_{\sigma}+\kappa^{a}{}_{b} R^{(0)}_{(-)}{}^{b}{}_{a})\right)
  -\frac{\alpha'}{4}\mathcal{D}^{(0)}_{(-)}\sigma^{a}{}_{b}\wedge
  \Omega^{(0)}_{(-)}{}^{b}{}_{a}\, ,
\end{equation}

\noindent
where the vertical Lorentz momentum map $\Pi_{\sigma}$ is such that, when
$\sigma^{a}{}_{b}=\kappa^{a}{}_{b}$

\begin{equation}
d\Pi_{\kappa} = \kappa^{a}{}_{b}R^{(0)}_{(-)}{}^{b}{}_{a}\, .  
\end{equation}

The conserved charge is the integral of the $(d-2)$-form
Eq.~(\ref{eq:Qsigmad-2}) for vertical Killing vector fields
$\kappa^{a}{}_{b}$ satisfying Eqs.~(\ref{eq:kappae=0}) and
(\ref{eq:Dkappa=0}). The first condition annihilates the first term,
corresponding to the Einstein-Hilbert term in the action but the rest
of the terms survive in this case and we get the non-vanishing Lorentz
charge

\begin{equation}
  \label{eq:Lorentzcharge}
  \mathcal{Q}[\kappa]
  =
  \frac{g_{s}^{(d)\, 2}}{16 \pi G_{N}^{(d)}}
  \int_{S^{d-2}}
  \left\{
        (-1)^{d}\frac{\alpha'}{2}\left[e^{-2\phi} \star d\Pi_{\kappa}
        + (-1)^{d}\Pi_{\kappa}\wedge \left(e^{-2\phi} \star H^{(0)}\right)\right]
    \right\}\, .
\end{equation}

In the proof of the first law we will find the integral of
$(d-2)$-form Eq.~(\ref{eq:Qsigmad-2}) for a Lorentz parameter that
satisfies Eq.~(\ref{eq:Dkappa=0}) only. This integral give, precisely,
the entropy.

\subsection{The  transformations under diffeomorphisms}
\label{sec-diffeos}

Now we turn our attention to the diffeomorphisms. Our treatment is
similar to the treatment of the $\delta_{\chi}$ gauge transformations,
although the use of compensating gauge transformations admits a more
general justification in terms of the gauge covariance of the modified
transformations (covariant Lie derivatives). Since we have discussed
at length these modifications in Refs.~\cite{Elgood:2020svt,Elgood:2020mdx}
we will only discuss the aspects not covered there: torsionful spin
connections, non-Abelian gauge fields and the more complicated
transformations of the KR 2-form.

In this section $k$ will always be a (horizontal) Killing vector which
generates a symmetry of the complete field configuration.

\subsubsection{Lie-Lorentz derivatives}
\label{sec-LieLorentz}

The transformations of the Vielbeins, the Levi-Civita spin connection and its
curvature 2-form have been discussed in Refs.~\cite{Elgood:2020svt,Elgood:2020mdx},
but it is convenient to adapt some of the formulae to the torsionful spin
connection. They are generically given in terms of the Lie-Lorentz (or
Lorentz-covariant Lie derivative
Refs.~\cite{kn:Lich,kn:Kos,kn:Kos2,Hurley:cf,Ortin:2002qb,Ortin:2015hya}) by
$\delta_{\xi} = -\mathbb{L}_{\xi}$. Therefore, we will continue this
discussion in terms of the latter.

The parameter of the compensating local Lorentz transformation that appears in
the Lie-Lorentz derivative of $\Omega^{(0)\, ab}_{(-)}$ is still given by

\begin{equation}
  \label{eq:LLparameter}
  \sigma_{\xi}{}^{ab}
  =
  \imath_{\xi}\omega^{ab} -\nabla^{[a}\xi^{b]}\, ,
\end{equation}

\noindent
but it is useful to rewrite it using
$\Omega^{(0)\, ab}_{(-)}$ in the covariant derivatives. Due to the
complete antisymmetry of the torsion, it takes the simple form

\begin{equation}
  \label{eq:LLparametertorsion}
  \sigma_{\xi}{}^{ab}
  =
  \imath_{\xi}\Omega^{(0)\, ab}_{(-)} -\mathcal{D}^{(0)}_{(+)}{}^{[a}\xi^{b]}\, . 
\end{equation}

\noindent
Observe that the presence of fully antisymmetric torsion does not modify the
Killing equation\footnote{The presence of generic torsion does modify the
  Killing equation.}

\begin{equation}
  \label{eq:Killingtorsion}
  2\mathcal{D}^{(0)}_{(\pm)\, (a}\xi_{b)} = 0\, .
\end{equation}

Notice that Eqs.~(\ref{eq:LLparametertorsion}) and (\ref{eq:Killingtorsion})
are completely independent of $H^{(0)}$ even if we have formally rewritten
them in terms of the torsionful spin connection $\Omega^{(0)}_{(-)}$.

The Lie-Lorentz derivative of the torsion $\imath_{b}\imath_{a}H^{(0)}$
follows the general formula while that of the Levi-Civita connection
$\omega^{ab}$ is given by

\begin{equation}
  \label{eq:LLomega}
  \mathbb{L}_{\xi}\omega^{ab}
  =
  \pounds_{\xi}\omega^{ab}-\mathcal{D}\sigma_{\xi}{}^{ab}\, ,
\end{equation}

\noindent
and, therefore, it is easy to see that

\begin{equation}
  \label{eq:LLOmega}
  \mathbb{L}_{\xi}\Omega^{(0)\, ab}_{(-)}
  =
  \pounds_{\xi}\Omega^{(0)\, ab}_{(-)}
  -\mathcal{D}^{(0)}_{(-)}\sigma_{\xi}{}^{ab}\, ,
\end{equation}

\noindent
and it is equally easy to see that it can be rewritten in the form

\begin{equation}
 \label{eq:LLOmega2}
  \mathbb{L}_{\xi}\Omega^{(0)\, ab}_{(-)}
  =
  \imath_{\xi}R^{(0)\, ab}_{(-)}
  +\mathcal{D}_{(-)}P_{(-)\xi}{}^{ab}\, ,
\end{equation}

\noindent
with

\begin{equation}
  \label{eq:P-xiab}
  P_{(-)\xi}{}^{ab}
  \equiv
  \mathcal{D}^{(0)}_{(+)}{}^{[a}\xi^{b]}\, ,
\end{equation}

\noindent
The identity

\begin{equation}
  \label{eq:Pab1(-)}
\xi^{\nu}R^{(0)}_{(-)\, \nu\mu}{}^{ab}
+\mathcal{D}^{(0)}_{(-)\, \mu}P_{(-)\xi}{}^{ab}
=
\mathcal{D}^{(0)}_{(-)}{}^{[a}\left(\nabla^{b]}\xi_{\mu}+\nabla_{\mu}\xi^{b]}\right)
-\tfrac{3}{2}\nabla_{[\mu|}\left(\xi^{\nu}H^{(0)}_{\nu|\rho\sigma]}\right)e^{a\rho}e^{b\sigma}\, ,
\end{equation}

\noindent
proves that
$\delta_{\xi}\Omega^{(0)\, ab}_{(-)} = -\mathbb{L}_{\xi}\Omega^{(0)\,
  ab}_{(-)}$ vanishes when $\xi^{\mu}=k^{\mu}$, because, in that case, 

\begin{equation}
  \label{eq:Pab-}
  -\imath_{k}R^{(0)\, ab}_{(-)}
  =
  \mathcal{D}^{(0)}_{(-)}P_{(-)k}{}^{ab}\, .
\end{equation}

\noindent
Because $P_{(-)k}{}^{ab}$ satisfies this equation, we will call it the
\textit{horizontal Lorentz momentum map associated to the torsionful
  spin connection}.

$k$, then, generates a diffeomorphism that leaves invariant the metric
and the KR 3-form field strength.

Again, $P_{(-)\xi}{}^{ab}$ is a Lorentz tensor and
$\delta_{\xi} \Omega^{(0)\, ab}_{(-)} = -\mathbb{L}_{\xi}\Omega^{(0)\,
  ab}_{(-)}$ is a Lorentz tensor although $\Omega^{(0)\, ab}_{(-)}$ is
a connection. When it vanishes, it vanishes in all Lorentz frames.

\subsubsection{Lie-Yang-Mills derivatives}
\label{sec-LieYangMills}

Since the spin connection is just the connection of the Lorentz group,
this case is very similar to the previous one, the main difference
being that the YM fields are fundamental fields while the spin
connection is a composite field. Apart from this, in many (but not
all, because of the absence of a YM analogue of the Vielbein)
instances we may just apply the same formulae with the sole change of
the adjoint group indices, as we are going to see.

In order to find the gauge-covariant Lie derivative of YM fields it is
convenient to consider the Lie-Lorentz derivative of the curvature tensor
first. In this case, since we do not know the form of the parameter of the
compensating gauge transformation, we can simply consider the standard Lie
derivative of the gauge field strength 2-form defined in Eq.~(\ref{eq:FAmn}):

\begin{equation}
  \pounds_{\xi}F^{A}
  =
  (\imath_{\xi}d +d\imath_{\xi})F^{A}
  =
  \mathcal{D}\imath_{\xi}F^{A}
  -f_{BC}{}^{A}\imath_{\xi}A^{B}F^{C}\, ,
\end{equation}

\noindent
where we have used the Bianchi identity $\mathcal{D}F^{A}=0$.

When $\xi=k$ this expression should vanish up to an infinitesimal
gauge transformation with some parameter that we denote by
$\tilde{\chi}_{k}{}^{A}$. Then,

\begin{equation}
  \label{eq:esa}
  \mathcal{D}\imath_{\xi}F^{A}
  =
  f_{BC}{}^{A}\left(\imath_{\xi}A^{B}+\tilde{\chi}_{k}{}^{B}\right)F^{C}
  \equiv
  f_{BC}{}^{A}P_{k}{}^{B}F^{C}\, ,
\end{equation}

\noindent
which, upon use of the Ricci identity Eq.~(\ref{eq:YMRicci}), can be
solved by a $P_{k}{}^{A}$ that we call the \textit{(horizontal)
  Yang-Mills momentum map} satisfying the equation

\begin{equation}
  \label{eq:PA}
  -\imath_{k}F^{A}
  =
  \mathcal{D}P_{k}{}^{A}\, .
\end{equation}

Eq.~(\ref{eq:Pab-}) is nothing by a particular case of this equation
for which the momentum map is explicitly known. This happens because
we know how to express the gauge field in terms of a more fundamental
field (the Vielbein).  In general, the general form of $P_{k}{}^{A}$
is not known but is determined up to a covariantly-constant gauge
parameter. We will use a $P_{\xi}{}^{A}$ which is undetermined except
for the fact that it reduces to $P_{k}{}^{A}$ satisfying
Eq.~(\ref{eq:PA}) for Killing vectors.

Now, we can use as definition of the Lie-Yang-Mills derivative of
$F^{A}$ the following expression which is guaranteed to vanish when
$\xi=k$ on account of Eq.~(\ref{eq:esa}):

\begin{equation}
    \mathbb{L}_{\xi}  F^{A}
    =
    \mathcal{D}\imath_{\xi}F^{A}
    -f_{BC}{}^{A}P_{\xi}{}^{B}F^{C}
 =
    \pounds_{\xi}F^{A} -\delta_{\chi_{\xi}}F^{A}\, ,
\end{equation}

\noindent
where the gauge compensating parameter $\chi_{\xi}{}^{A}$ is given by
the (now usual) expression

\begin{equation}
  \label{eq:chixiA}
  \chi_{\xi}{}^{A}
  =
  \imath_{\xi}A^{A}-P_{\xi}{}^{A}\, .
\end{equation}

The Lie-Yang-Mills derivative of the gauge field is, then

\begin{equation}
  \label{eq:LLA}
    \mathbb{L}_{\xi}A^{A}
     \equiv
    \pounds_{\xi}A^{A} -\mathcal{D}\chi_{\xi}{}^{A}
 =
    \imath_{\xi}F^{A} +\mathcal{D}P_{\xi}{}^{A}\, ,
\end{equation}

\noindent
and, by construction, it vanishes automatically when $\xi$ is a Killing vector
field $k^{\mu}$ and $P_{k}{}^{A}$ is the momentum map satisfying
Eq.~(\ref{eq:PA}).

\subsubsection{The Kalb-Ramond field}
\label{sec-firstorderkalbdramond}

The parameters of the compensating YM and local Lorentz
transformations of the KR field are the same transformations
$\chi_{\xi}{}^{A}$ and $\sigma_{\xi}{}^{ab}$ that we perform on other
fields with YM and Lorentz indices, given by Eqs.~(\ref{eq:chixiA})
and (\ref{eq:LLparameter}).  Thus, if we want to construct a
transformation of this field under diffeomorphisms that annihilates it
when $\xi=k$ by combining its standard Lie derivative with gauge
transformations, the only gauge parameter we can still play with is
the 1-form $\Lambda$ because the rest are already completely
determined. We have

\begin{equation}
  \begin{aligned}
    \delta_{\xi} B
    = &
    -\pounds_{\xi}B
    +(\delta_{\Lambda_{\xi}}+\delta_{\chi_{\xi}}+\delta_{\sigma_{\xi}})
    B
    \\
    & \\
    = &
    -\pounds_{\xi}B
    +d\Lambda_{\xi}
    -\frac{\alpha'}{4}\chi_{\xi\, A}dA^{A}
    -\frac{\alpha'}{4}\sigma_{\xi}{}^{a}{}_{b}
    d\Omega^{(0)\, b}_{(-)}{}_{a}\, .
  \end{aligned}
\end{equation}

Again, it is convenient to start by considering the transformation of the
3-form field strength $H^{(1)}$ defined in Eq.~(\ref{eq:H1def}) under
diffeomorphisms, because it is gauge invariant:

\begin{equation}
  \begin{aligned}
    \delta_{\xi}H^{(1)}
    = &
    -\pounds_{\xi}H^{(1)}
    \\
    & \\
    = & -\imath_{\xi}dH^{(1)} -d\imath_{\xi}H^{(1)}
    \\
    & \\
    = & 
    -d\imath_{\xi}H^{(1)}
    -\frac{\alpha'}{2}\left(\imath_{\xi}F_{A}\wedge F^{A}
    +\imath_{\xi}R^{(0)}_{(-)}{}^{a}{}_{b}\wedge R^{(0)}_{(-)}{}^{b}{}_{a} \right)\, ,
\end{aligned}
\end{equation}

\noindent
where we have used the Bianchi identity Eq.~(\ref{eq:BianchiH1}).

When $\xi=k$ we can use Eqs.~(\ref{eq:Pab-}) and (\ref{eq:PA}),
integrate by parts, and use now the Bianchi identities for the
curvatures, getting:

\begin{equation}
  \begin{aligned}
    \delta_{k}H^{(1)}
    = &
    -d\imath_{k}H^{(1)}
    +\frac{\alpha'}{2}\left(\mathcal{D}P_{k\, A}\wedge F^{A}
    +\mathcal{D}_{(-)}P_{(-)\,k}{}^{a}{}_{b}\wedge R^{(0)}_{(-)}{}^{b}{}_{a}
  \right)
  \\
  & \\
  = & 
    -d\left[\imath_{k}H^{(1)}-\frac{\alpha'}{2}\left(P_{k\, A} F^{A}
    +P_{(-)\,k}{}^{a}{}_{b} R^{(0)}_{(-)}{}^{b}{}_{a}\right)\right]\, .
\end{aligned}
\end{equation}

\noindent
By assumption, the above expression must vanish
identically. Therefore, locally, there must exist a gauge-invariant
1-form, the \textit{horizontal Kalb-Ramond momentum map} $P_{k}$,
satisfying

\begin{equation}
  \label{eq:Pkmu1alpha}
  -\imath_{k}H^{(1)}+\frac{\alpha'}{2}\left(P_{k\, A} F^{A}
    +P_{(-)\,k}{}^{a}{}_{b} R^{(0)}_{(-)}{}^{b}{}_{a}\right)
  =
  dP_{k}\, .
\end{equation}

\noindent
Then, if we apply the rule of thumb that the parameter of the
compensating gauge transformation is the inner product of the vector
that generates the diffeomorphisms with the ``connection'' (here $B$)
minus the momentum map (here some 1-form $P_{\xi}$ that in this case
satisfies Eq.~(\ref{eq:Pkmu1alpha}) when $\xi=k$)

\begin{equation}
  \label{eq:lambdaxim}
  \Lambda_{\xi}
  =
   \imath_{\xi}B-P_{\xi}\, ,
\end{equation}

\noindent
we arrive at the following candidate to $\delta_{\xi}B$:

\begin{equation}
  \begin{aligned}
    \delta_{\xi} B
     = &
    -\pounds_{\xi}B
    +d\Lambda_{\xi}
    -\frac{\alpha'}{4}\left(\chi_{\xi\, A}dA^{A}+\sigma_{\xi}{}^{a}{}_{b}
    d\Omega^{(0)\, b}_{(-)}{}_{a}\right)
    \\
    & \\ 
    = &
    -\imath_{\xi}H^{(1)}
      -\frac{\alpha'}{4}\left(A_{A}\wedge \imath_{\xi}F^{A}
      +\Omega^{(0)\, a}_{(-)}{}_{b}\wedge  \imath_{\xi}R^{(0)\, b}_{(-)}{}_{a}
      \right)
    \\
    & \\
    &
    -dP_{\xi}
    +\frac{\alpha'}{4}\left(P_{\xi\, A}dA^{A}
    +P_{(-)\, \xi}{}^{a}{}_{b}d\Omega^{(0)\, b}_{(-)}{}_{a}\right)\, .
\end{aligned}
\end{equation}

Let us see if, with this definition, $\delta_{k}B=0$. Using
Eqs.~(\ref{eq:Pkmu1alpha}), (\ref{eq:PA}) and (\ref{eq:Pab-}) we get, instead
of zero, a total derivative

\begin{equation}
  \begin{aligned}
    \delta_{k} B
    = &
    -\frac{\alpha'}{4}d\left(P_{k\, A}A^{A}
      +P_{(-)\, k}{}^{a}{}_{b}\Omega^{(0)\, b}_{(-)}{}_{a}\right)\, ,
\end{aligned}
\end{equation}

\noindent
which we can simple absorb in redefinition of $\Lambda_{\xi}$ in
Eq.~(\ref{eq:lambdaxim}):

\begin{equation}
  \label{eq:lambdaxim2}
    \Lambda_{\xi}
\equiv
   \imath_{\xi}B-P_{\xi}
    +\frac{\alpha'}{4}d\left(P_{\xi\, A}A^{A}
      +P_{(-)\, \xi}{}^{a}{}_{b}\Omega^{(0)\, b}_{(-)}{}_{a}\right)\, .
\end{equation}

\noindent
With this new parameter,

\begin{equation}
  \label{eq:deltaB1}
  \begin{aligned}
  \delta_{\xi} B
  = &
      -\pounds_{\xi}B
    +d\Lambda_{\xi}
    -\frac{\alpha'}{4}\chi_{\xi\, A}dA^{A}
    -\frac{\alpha'}{4}\sigma_{\xi}{}^{a}{}_{b}
    d\Omega^{(0)\, b}_{(-)}{}_{a}
  \\
  & \\
  = &
  -\left[\imath_{\xi}H^{(1)}
    -\frac{\alpha'}{2}\left(P_{\xi\, A}F^{A}
      +P_{(-)\, \xi}{}^{a}{}_{b}R^{(0)\, b}_{(-)}{}_{a}\right)
    +dP_{k}
  \right]
  \\
  & \\
  & \hspace{.5cm}
      +\frac{\alpha'}{4}\left(A_{A}\wedge \delta_{\xi}A^{A}
      +\Omega^{(0)\, a}_{(-)}{}_{b}\wedge  \delta_{\xi}\Omega^{(0)\, b}_{(-)}{}_{a}
    \right)
    \\
    & \\
    & \equiv
    -\mathbb{L}_{\xi}B\, ,
\end{aligned}
\end{equation}

\noindent
that vanishes identically when $\xi=k$ by virtue of the definition of
the KR momentum map Eq.~(\ref{eq:Pkmu1alpha}) and of
$\delta_{\xi}A^{A} =\delta_{\xi}\Omega^{(0)\, b}_{(-)}{}_{a}=0$.

The behavior of this variation under gauge transformations is far from
obvious. A direct calculation gives

\begin{equation}
  \delta_{\rm gauge} \delta_{\xi} B
  =
  \frac{\alpha'}{4}
  \left(
    d\chi_{A}\wedge \delta_{\xi}A^{A}
    +d\sigma^{a}{}_{b}\wedge \delta_{\xi}\Omega^{(0)\, b}_{(-)}{}_{a}
  \right)\, ,
\end{equation}

\noindent
with $\delta_{\xi}A^{A}= -\mathbb{L}_{\xi}A^{A}$ with the Lie-Yang-Mills
covariant derivative given by Eq.~(\ref{eq:LLA}) and with
$\delta_{\xi}\Omega^{(0)\, ab}_{(-)} = -\mathbb{L}_{\xi}\Omega^{(0)\,
  ab}_{(-)}$, with the Lie-Lorentz derivative given by
Eq.~(\ref{eq:LLOmega2}). Therefore, although the $\delta_{\xi} B$ defined
above is not gauge-invariant, $\delta_{k} B$ vanishes in a gauge-invariant
way.

\subsection{The Wald-Noether charge}
\label{sec-firstordervariationactiondiffeos}

Now we consider the variation of the action $S^{(1)}$ given in
Eq.~(\ref{eq:heterotic1diff}) under the transformations
$\delta_{\xi}=-\mathbb{L}_{\xi}$ for all the fields, where $\mathbb{L}_{\xi}$
is the gauge-covariant derivative which, for the Vielbein is given by
\cite{Elgood:2020svt}

\begin{equation}
      \label{eq:LLeam2}
  \mathbb{L}_{\xi}e^{a}
 =
\mathcal{D}\xi^{a}+P_{\xi}{}^{a}{}_{b}e^{b}\, ,
\end{equation}

\noindent
for the torsionful spin connection in Eq.~(\ref{eq:LLOmega2}), for the YM
fields in Eq.~(\ref{eq:LLA}) and for the KR field in Eq.~(\ref{eq:deltaB1}).

From Eq.~(\ref{eq:deltaS1})

\begin{equation}
  \label{eq:deltaxiS1}
  \begin{aligned}
    \delta_{\xi} S^{(1)}
    & =
    -\int \left\{ \mathbf{E}^{(1)}_{{\rm exp}\, a}\wedge
      \left(\mathcal{D}\imath_{\xi}e^{a}+P_{\xi}{}^{a}{}_{b}e^{b}\right)
\right.      
      +\mathbf{E}^{(1)}_{\phi}\imath_{\xi}d\phi
            \\
      & \\
      &
+\mathbf{E}^{(1)}_{A} \wedge \left(\imath_{\xi}F^{A} +\mathcal{D}P_{\xi}{}^{A} \right)
      +\mathbf{E}^{(1)\, b}{}_{a}\wedge
\left(\imath_{\xi}R^{(0)\, a}_{(-)}{}_{b}+\mathcal{D}_{(-)}P_{(-)\xi}{}^{a}{}_{b} \right)
      \\
      & \\
      &
      +\mathbf{E}^{(1)}_{{\rm exp}\, B}\wedge \left[
            \imath_{\xi}H^{(1)}
      +\frac{\alpha'}{4}\left(A_{A}\wedge \imath_{\xi}F^{A}
      +\Omega^{(0)\, a}_{(-)}{}_{b}\wedge  \imath_{\xi}R^{(0)\, b}_{(-)}{}_{a}
      \right)
        \right.
      \\
      & \\
      &
    -\frac{\alpha'}{4}\left(P_{\xi\, A}dA^{A}
    +P_{(-)\, \xi}{}^{a}{}_{b}d\Omega^{(0)\, b}_{(-)}{}_{a}\right)
      \left.
        +d\left[P_{\xi}
-\frac{\alpha'}{4}\left(P_{\xi\, A}A^{A}
      +P_{(-)\, \xi}{}^{a}{}_{b}\Omega^{(0)\, b}_{(-)}{}_{a}\right)
  \right]
  \right]
    \\
    & \\
    &
    \hspace{.5cm}
    \left.
      -d\mathbf{\Theta}^{(1)}(\varphi,\delta_{\xi}\varphi) \right\}\, ,
\end{aligned}
\end{equation}

\noindent
where $\mathbf{\Theta}^{(1)}(\varphi,\delta_{\xi}\varphi)$ is given by

\begin{equation}
  \label{eq:Theta1xi}
\begin{aligned}
    \mathbf{\Theta}^{(1)}(\varphi,\delta_{\xi}\varphi)
    = &
     e^{-2\phi}\star (e^{a}\wedge e^{b})\wedge
    \left(  \imath_{\xi}R_{ab} +\mathcal{D}P_{\xi\, ab}\right)
    -2\imath_{a}de^{-2\phi}\star(e^{a}\wedge e^{b})\wedge
    \left(\mathcal{D}\imath_{\xi}e_{b}+P_{\xi\, bc}e^{c}\right)
    \\
    & \\
    &
    +8e^{-2\phi}\star d\phi \imath_{\xi}d\phi
        \\
    & \\
    &
-e^{-2\phi}\star H^{(1)}\wedge \left\{   \imath_{\xi}H^{(1)}
      +\frac{\alpha'}{4}\left(A_{A}\wedge \imath_{\xi}F^{A}
      +\Omega^{(0)\, a}_{(-)}{}_{b}\wedge  \imath_{\xi}R^{(0)\, b}_{(-)}{}_{a}
      \right)
        \right.
      \\
      & \\
      &
    -\frac{\alpha'}{4}\left(P_{\xi\, A}dA^{A}
    +P_{(-)\, \xi}{}^{a}{}_{b}d\Omega^{(0)\, b}_{(-)}{}_{a}\right)
      \left.
        +d\left[P_{\xi}
-\frac{\alpha'}{4}\left(P_{\xi\, A}A^{A}
      +P_{(-)\, \xi}{}^{a}{}_{b}\Omega^{(0)\, b}_{(-)}{}_{a}\right)
  \right]
  \right\}
    \\
    & \\
    &
-\frac{\alpha'}{2}e^{-2\phi}\left(\star F_{A} -\tfrac{1}{2}\star H^{(0)}\wedge
  A_{A}\right)\wedge \left(\imath_{\xi}F^{A} +\mathcal{D}P_{\xi}{}^{A} \right)\, .
    \\
    & \\
    &
-\frac{\alpha'}{2}e^{-2\phi}\left(\star R^{(0)}_{(-)}{}^{b}{}_{a}
      -\tfrac{1}{2}\star H^{(0)}\wedge
      \Omega^{(0)}_{(-)}{}^{b}{}_{a}\right)\wedge
    \left(\imath_{\xi}R^{(0)\, a}_{(-)}{}_{b}+\mathcal{D}_{(-)}P_{(-)\xi}{}^{a}{}_{b} \right)\, .
  \end{aligned}
\end{equation}

Integrating by parts and using the Noether identities
Eqs.~(\ref{eq:noetheridentity1}), (\ref{eq:noetheridentity2}),
(\ref{eq:noetheridentity3}), (\ref{eq:noetheridentity5}) and the Noether
identity associated to the invariance under diffeomorphisms

\begin{equation}
  \label{eq:noetherdifffirst}
  \begin{aligned}
    & (-1)^{d}\mathcal{D}\mathbf{E}^{(1)}_{{\rm exp}\, a}\imath_{\xi}e^{a}
    +\mathbf{E}^{(1)}_{{\rm exp}\, B}\wedge \imath_{\xi}H^{(1)}
    +\mathbf{E}^{(1)}_{\phi}\imath_{\xi}d\phi 
  \\
  & \\
  & +\left(\mathbf{E}^{(1)}_{A} +\frac{\alpha'}{4}\mathbf{E}^{(0)}_{{\rm
        exp}\, B}\wedge A_{A}\right)\wedge \imath_{\xi}F^{A}
  +\left(\mathbf{E}^{(1)\, b}{}_{a} +\frac{\alpha'}{4}\mathbf{E}^{(0)}_{{\rm
        exp}\, B}\wedge \Omega^{(0)\, b}_{(-)}{}_{a} \right)\wedge
  \imath_{\xi}R^{(0)\, a}_{(-)}{}_{b}
  \\
  & \\
  & = 0\, ,
\end{aligned}
\end{equation}

\noindent
we can see that the volume term in the variation of the action
Eq.~(\ref{eq:deltaxiS1}) reduces to another total derivative

\begin{equation}
  \label{eq:deltaxiS3}
    \delta_{\xi} S^{(1)}
     =
    \int d\mathbf{\Theta}^{(1)\, \prime}(\varphi,\delta_{\xi}\varphi)\, ,
\end{equation}

\noindent
with

\begin{equation}
  \label{eq:Thetaprime}
  \begin{aligned}
    \mathbf{\Theta}^{(1)\, \prime}(\varphi,\delta_{\xi}\varphi)
    =\,\, &
    \mathbf{\Theta}^{(1)}(\varphi,\delta_{\xi}\varphi)
    \\
    & \\
    &
    + (-1)^{d}\mathbf{E}^{(1)}_{{\rm exp}\,
      a}\imath_{\xi}e^{a} +(-1)^{d-1}\mathbf{E}^{(1)}_{{\rm exp}\, B}\wedge P_{\xi}
  \\
  & \\
  & +(-1)^{d}\left(\mathbf{E}^{(1)}_{A}
      +\frac{\alpha'}{4}\mathbf{E}^{(0)}_{{\rm exp}\, B}\wedge
      A_{A}\right)P_{\xi}{}^{A}
      \\
  & \\
  & +(-1)^{d}\left(\mathbf{E}^{(1)\, b}{}_{a}
      +\frac{\alpha'}{4}\mathbf{E}^{(0)}_{{\rm exp}\, B}\wedge \Omega^{(0)\,
        b}_{(-)}{}_{a} \right) P_{(-)\xi}{}^{a}{}_{b}\, .
\end{aligned}
\end{equation}

The usual reasoning leads us to the off-shell identity

\begin{equation}
  \label{eq:dJ1=0}
  d\mathbf{J}^{(1)}[\xi]=0\, ,
\end{equation}

\noindent
where

\begin{equation}
  \mathbf{J}^{(1)}[\xi]
  \equiv
d\mathbf{\Theta}^{(1)\, \prime}(\varphi,\delta_{\xi}\varphi)
  +\imath_{\xi}\mathbf{L}^{(1)}\, ,
\end{equation}

\noindent
and to the local existence of a $(d-2)$-form $\mathbf{Q}^{(1)}[\xi]$
such that $ \mathbf{J}^{(1)}[\xi] = d\mathbf{Q}^{(1)}[\xi]$.

A straightforward calculation leads to the fully gauge-invariant
Wald-Noether charge 

\begin{equation}
  \label{eq:Q1}
\begin{aligned}
\mathbf{Q}^{(1)}[\xi]
    = &
    (-1)^{d}\star (e^{a}\wedge e^{b})
    \left[e^{-2\phi}P_{\xi\, ab}-2\imath_{a}de^{-2\phi}\xi_{b}\right]
  \\
  & \\
  &
    +(-1)^{d-1}\frac{\alpha'}{2}
    \left[P_{\xi\, A}e^{-2\phi}\star F^{A}
      +P_{(-)\xi}{}^{a}{}_{b}\left(e^{-2\phi}\star R^{(0)\, b}_{(-)}{}_{a}\right)\right]
  \\
  & \\
  &
      -P_{\xi}\wedge\left(e^{-2\phi}\star H^{(1)}\right)\, ,
  \end{aligned}
\end{equation}

\noindent
which is one of the main results of this paper.

\section{Restricted generalized zeroth laws}
\label{sec-zerothlaws}

One of the main ingredients in Wald's approach to the first law of
black hole mechanics is the zeroth law stating that $\kappa$ is
constant over the horizon \cite{Bardeen:1973gs}. Originally, this law
was proved using the Einstein equations and the dominant energy
condition (see, for instance, Ref.~\cite{Wald:1984rg}) but a
completely geometrical proof was presented in Ref.~\cite{Racz:1995nh}.

In presence of an electromagnetic field one also needs to use the
\textit{generalized zeroth law} that guarantees that the electrostatic
potential is also constant over the whole horizon. There is no purely
geometrical proof of this law, though, and the standard proof also
makes use of the Einstein equations and of the dominant energy
condition. In Ref.~\cite{Elgood:2020mdx} we have explained how this
proof can be extended to a theory containing an arbitrary number of
Abelian vector fields and the KR field coupled to them via
Chern-Simons terms. Essentially one gets a sum of non-negative terms
containing the contribution of each field, and each of them has to
vanish. Extending this proof to the non-Abelian case, as long as we
restrict ourselves to a gauge group with definite positive Killing
metric because one gets sums of non-negative terms. However, the
$R^{(0)\, 2}_{(-)}$ term of our theory is of YM type, but with
non-definite Killing metric because of the non-compactness of the
Lorentz group and the proof cannot be extended to this case in a
streightforward manner.

It is, however, possible to proof the first law in bifurcate horizons
if one can proof generalized zeroth laws for the matter fields
restricted to the bifurcation sphere $\mathcal{BH}$ where the Killing
vector associated to the event horizon, $k$, vanishes
identically. These \textit{restricted generalized zeroth laws} state
the closedness of certain differential forms on $\mathcal{BH}$. The
definitions of the potentials as certain constants follow from them as
we are going to explain.

Assuming all the fields are regular over the horizon, it is clear that the
inner products of their field strengths with $k$ must vanish on
$\mathcal{BH}$:

\begin{subequations}
  \begin{align}
    \label{eq:idk=0}
    \imath_{k}d\phi & \stackrel{\mathcal{BH}}{=} 0\, , \\
                    & \nonumber \\
    \label{eq:ikH=0}
    \imath_{k}H & \stackrel{\mathcal{BH}}{=} 0\, , \\
                    & \nonumber \\
    \label{eq:ikFA=0}
    \imath_{k}F^{A} & \stackrel{\mathcal{BH}}{=} 0\, , \\
                    & \nonumber \\
    \label{eq:ikR=0}
    \imath_{k}R^{(0)\, a}_{(-)}{}_{b} & \stackrel{\mathcal{BH}}{=} 0\, . \\
  \end{align}
\end{subequations}

Eq.~(\ref{eq:idk=0}) is actually true over the whole spacetime, by
assumption. From Eq.~(\ref{eq:ikFA=0}) and the definition of the YM momentum
map $P_{k}{}^{A}$ we find that

\begin{equation}
  \label{eq:PkABH0}
  \mathcal{D}P_{k}{}^{A}  \stackrel{\mathcal{BH}}{=} 0\, ,
\end{equation}

\noindent
which tells us that the horizontal YM momentum map $P_{k}{}^{A}$ is,
at the same time, a vertical Killing vector field on $\mathcal{BH}$.
This is what we need in order to have an associated conserved charge
there (see the discussion in Section~\ref{sec-YMcharge}).

Analogously, from Eq.~(\ref{eq:ikR=0}) and the definition of the momentum map
$P_{(-)k}{}^{a}{}_{b}$ Eq.~(\ref{eq:Pab-}) we get

\begin{equation}
  \label{eq:PkabBH0}
  \mathcal{D}^{(0)}_{(-)}P_{(-)k}{}^{a}{}_{b}  \stackrel{\mathcal{BH}}{=} 0\, ,
\end{equation}

\noindent
which tells us that the horizontal Lorentz momentum map $P_{k}{}^{A}$
is, also, a vertical Killing vector field on $\mathcal{BH}$.

Observe that the last two equations have as consequence the existence of the
gauge-invariant 1-forms $\Psi_{P_{k}}$ and $\Pi_{P_{k}}$ defined by 

\begin{subequations}
  \begin{align}
    \label{eq:PiPk}
  d\Pi_{P_{k}} 
  & \stackrel{\mathcal{BH}}{=}
     P_{(-)k}{}^{a}{}_{b}R^{(0)\, b}_{(-)}{}_{a}\, ,
    \hspace{1cm}
    \\
    & \nonumber \\
        \label{eq:PsiPk}
  d\Psi_{P_{k}}
  & \stackrel{\mathcal{BH}}{=}
  P_{k\, A}F^{A}\, .
  \end{align}
\end{subequations}

\noindent
The closedness of the right-hand sides of these equations on
$\mathcal{BH}$, which guarantee the local existence of $\Psi_{P_{k}}$
and $\Pi_{P_{k}}$ there are the restricted generalized zeroth laws for
the YM and torsionful spin connecton fields.

Finally, from Eq.~(\ref{eq:ikH=0}) and the definition of the KR
momentum map Eq.~(\ref{eq:Pkmu1alpha}) plus the above two equations
that define $\Psi_{P_{k}}$ and $\Pi_{P_{k}}$ we get

\begin{equation}
  \label{eq:KRrestrictedgeneralizedzerothlaw}
d\left[P_{k}  -\frac{\alpha'}{2}\left(\Psi_{P_{k}}+\Pi_{P_{k}}\right)\right]
   \stackrel{\mathcal{BH}}{=}
0\, ,
\end{equation}

\noindent
which is the restricted generalized zeroth law of the KR field.

\section{The first law}
\label{sec-firstlaw}

Following Wald \cite{Wald:1993nt}, we start by defining the
\textit{pre-symplectic $(d-1)$-form} \cite{Lee:1990nz}

\begin{equation}
\omega^{(1)}(\varphi,\delta_{1}\varphi,\delta_{2} \varphi)  
\equiv 
\delta_{1}\mathbf{\Theta}^{(1)}(\varphi,\delta_{2} \varphi)
-\delta_{2}\mathbf{\Theta}^{(1)}(\varphi,\delta_{1} \varphi)\, ,
\end{equation}

\noindent
and the \textit{symplectic form} relative to the Cauchy surface $\Sigma$

\begin{equation}
\Omega^{(1)}(\varphi,\delta_{1}\varphi,\delta_{2} \varphi)  
\equiv
\int_{\Sigma}\omega^{(1)}(\varphi,\delta_{1}\varphi,\delta_{2} \varphi)\, .
\end{equation}

When $\varphi$ is a solution of the equations of motion
$\mathbf{E}_{\varphi}=0$, $\delta_{1}\varphi=\delta\varphi$ is an
arbitrary variation of the fields and
$\delta_{2}\varphi= \delta_{\xi}\varphi$ is their variation under
diffeomorphisms \cite{Iyer:1994ys}

\begin{equation}
  \omega^{(1)}(\varphi,\delta\varphi,\delta_{\xi}\varphi)
  =
  \delta\mathbf{J}^{(1)}+d\imath_{\xi}\mathbf{\Theta}^{(1)\, \prime}
  =
  \delta d\mathbf{Q}^{(1)}[\xi]+d\imath_{\xi}\mathbf{\Theta}^{(1)\, \prime}\, ,
\end{equation}

\noindent
where, in our case, the Noether-Wald $(d-2)$-form charge
$\mathbf{Q}^{(1)}$ is given by Eq.~(\ref{eq:Q1}) and
$\mathbf{\Theta}'$ is given in Eq.~(\ref{eq:Thetaprime}). Since,
on-shell, $\mathbf{\Theta}^{(1)} = \mathbf{\Theta}^{(1)\, \prime}$, we
have that, if $\delta\varphi$ satisfies the linearized equations of
motion, $\delta d\mathbf{Q}^{(1)}= d\delta \mathbf{Q}^{(1)}$.
Furthermore, if the parameter $\xi=k$ generates a transformation that
leaves invariant the field configuration,
$\delta_{k}\varphi=0$,\footnote{Notice that our goal in
  Section~\ref{sec-diffeos} was, precisely, to construct variations of
  the fields $\delta_{\xi}$ with that property.} linearity implies
that $\omega^{(1)}(\varphi,\delta\varphi,\delta_{k}\varphi)=0$, and

\begin{equation}
d\left( \delta \mathbf{Q}^{(1)}[k]+\imath_{k}\mathbf{\Theta}^{(1)\, \prime}  \right)=0\, .
\end{equation}

\noindent
Integrating this expression over a hypersurface $\Sigma$ with boundary
$\delta\Sigma$ and using Stokes' theorem we arrive at

\begin{equation}
  \int_{\delta\Sigma}
  \left( \delta \mathbf{Q}^{(1)}[k]+\imath_{k}\mathbf{\Theta}^{(1)\, \prime}  \right)
  =
  0\, .
\end{equation}

We consider field configurations that describe asymptotically flat,
stationary, black-hole spacetimes with bifurcate horizons
$\mathcal{H}$ and the Killing vector $k$ is the one whose Killing
horizon is the black hole's event horizon. $k$, then, will be given by
a linear combination with constant coefficients $\Omega^{n}$ of the
timelike Killing vector associated to stationarity,
$t^{\mu}\partial_{\mu}$ and the $[\tfrac{1}{2}(d-1)]$ generators of
inequivalent rotations in $d$ spacetime dimensions
$\phi_{n}^{\mu}\partial_{\mu}$

\begin{equation}
  k^{\mu} = t^{\mu} +\Omega^{n}\phi_{n}^{\mu}\, .  
\end{equation}

\noindent
The constant coefficients $\Omega^{n}$ are the angular velocities of
the horizon.

The hypersurface $\Sigma$ to be the space bounded by infinity and the
bifurcation sphere $\mathcal{BH}$ on which $k=0$, so $\delta\Sigma$
has two disconnected pieces: a $(d-2)$-sphere at infinity,
S$^{d-2}_{\infty}$, and the bifurcation sphere $\mathcal{BH}$.  Then,
taking into account that $k=0$ on $\mathcal{BH}$, we obtain the
relation

\begin{equation}
  \delta \int_{\mathcal{BH}}   \mathbf{Q}^{(1)}[k]
  =
  \int_{\mathrm{S}^{d-2}_{\infty}}
  \left( \delta \mathbf{Q}^{(1)}[k]+\imath_{k}\mathbf{\Theta}^{(1)\ \prime} \right)\, .
\end{equation}

As explained in Ref.~\cite{Iyer:1994ys,Compere:2007vx}, the right-hand
side can be identified with $\delta M -\Omega^{m}\delta J_{n}$, where
$M$ is the total mass of the black-hole spacetime and $J_{n}$ are the
independent components of the angular momentum.\footnote{When the
  spacetime has compact dimensions, the $d$-dimensional mass $M$ is a
  combination of the lower-dimensional mass and Kaluza-Klein
  charges. The details depend on the compactification and will be
  studied elsewhere.}

Using the explicit form of $\mathbf{Q}^{(1)}[k]$, Eq.~(\ref{eq:Q1}),
noticing that
$-2\imath_{a}de^{-2\phi}k_{b}\stackrel{\mathcal{BH}}{=} 0$ and
restoring the overall factor
$g^{(d)\, 2}_{s}(16\pi G^{(d)}_{N})^{-1}$, we find

\begin{equation}
  \label{eq:deltaintQ1}
\begin{aligned}
 \delta \int_{\mathcal{BH}} \mathbf{Q}^{(1)}[k]
 & =
 \frac{g_{s}^{(d)\, 2}}{16\pi G_{N}^{(d)}}
 \int_{\mathcal{BH}}
     (-1)^{d}e^{-2\phi}\star (e^{a}\wedge e^{b}) P_{k\, ab}
  \\
  & \\
  & \hspace{.5cm}
        +\frac{g_{s}^{(d)\, 2}}{16\pi G_{N}^{(d)}}\int_{\mathcal{BH}}
(-1)^{d-1}\frac{\alpha'}{2}
      P_{(-)k}{}^{a}{}_{b}\left(e^{-2\phi}\star R^{(0)\, b}_{(-)}{}_{a}\right)
  \\
  & \\
  & \hspace{.5cm}
    +\frac{g_{s}^{(d)\, 2}}{16\pi G_{N}^{(d)}}\int_{\mathcal{BH}}
(-1)^{d-1}\frac{\alpha'}{2}
    P_{k\, A}e^{-2\phi}\star F^{A}
  \\
  & \\
  & \hspace{.5cm}
      - \frac{g_{s}^{(d)\, 2}}{16\pi G_{N}^{(d)}}
 \int_{\mathcal{BH}}
P_{k}\wedge\left(e^{-2\phi}\star H^{(1)}\right)\, .
\end{aligned}
\end{equation}

The right-hand side ot this identity is expected to be of the form
$T\delta S+\Phi\delta\mathcal{Q}$ for some charges $\mathcal{Q}$ and
potentials $\Phi$. However, when we compare the third and fourth
integrals in the right-hand side with the definitions of the YM and KR
charges Eqs.~(\ref{eq:YMcharge}) and (\ref{eq:QLambdacharge}) we see
that some terms are missing in the integrand of the first and that, in
the second, there is no closed or harmonic form in the integrand,
since the horizontal KR momentum map is not necessarily closed on
$\mathcal{BH}$. We found a similar problem in
Ref.~\cite{Elgood:2020mdx} and the solution is essentially the same:
add and subtract the same term in different integrals in order to
complete the integrand of the definition of YM charge and in order to
construct a 1-form which is closed in $\mathcal{BH}$.

The 1-form shich is closed on $\mathcal{BH}$ and which contains
$P_{k}$ follows from the restricted generalized zeroth law of the KR
field, Eq.~(\ref{eq:KRrestrictedgeneralizedzerothlaw}). We must add a
term $-\frac{\alpha'}{2}\Psi_{P_{k}}$ to the fourth integral and
substract the same term to the third, which now contains all the terms
associated to the YM charge becuase of the restricted generalized
zeroth law Eq.~(\ref{eq:PkABH0}). However,
Eq.~(\ref{eq:KRrestrictedgeneralizedzerothlaw}) also tells us to add
another term $-\frac{\alpha'}{2}\Pi_{P_{k}}$ to the fourth integral
and we can only compensate by subtracting it to the second. This
completes the closed 1-form in the fourth integral and completes the
integrand of the Lorentz charge according to
Eq.~(\ref{eq:Lorentzcharge}) and thanks to the restricted generalized
zeroth law Eq.~(\ref{eq:PkabBH0}).

The result of these additions and subtractions is

\begin{equation}
  \label{eq:deltaintQ1-2}
\begin{aligned}
 \delta \int_{\mathcal{BH}} \mathbf{Q}^{(1)}[k]
 & =
 \frac{g_{s}^{(d)\, 2}}{16\pi G_{N}^{(d)}}
 \int_{\mathcal{BH}}
     (-1)^{d}e^{-2\phi}\star (e^{a}\wedge e^{b})
    P_{k\, ab}
  \\
  & \\
  & \hspace{.5cm}
        +\frac{g_{s}^{(d)\, 2}}{16\pi G_{N}^{(d)}}\int_{\mathcal{BH}}
        (-1)^{d-1}\frac{\alpha'}{2}\left[e^{-2\phi}\star d\Pi_{P_{k}}
          +(-1)^{d}\Pi_{P_{k}}\wedge\left(e^{-2\phi}\star H^{(0)}\right)\right]
  \\
  & \\
  & \hspace{.5cm}
    +\frac{g_{s}^{(d)\, 2}}{16\pi G_{N}^{(d)}}\int_{\mathcal{BH}}
(-1)^{d-1}\frac{\alpha'}{2}\left[
  e^{-2\phi}\star d\Psi_{P_{k}}
  +(-1)^{d}\Psi_{P_{k}}\wedge\left(e^{-2\phi}\star H^{(0)}\right)\right]
        \\
  & \\
  & \hspace{.5cm}
      - \frac{g_{s}^{(d)\, 2}}{16\pi G_{N}^{(d)}}
 \int_{\mathcal{BH}}
 \left[P_{k}
   -\frac{\alpha'}{2}(\Psi_{P_{k}}+\Pi_{P_{k}})\right]\wedge
 \left(e^{-2\phi}\star H^{(1)}\right)\, .
\end{aligned}
\end{equation}

\noindent
where $\Psi_{P_{k}}$ and $\Pi_{P_{k}}$ satisfy Eqs.~(\ref{eq:PsiPk})
and (\ref{eq:PiPk}), respectively, whose integrability is guaranteed
by the fact that the YM and Lorentz momentum maps are covariantly
constant on $\mathcal{BH}$ (the restricted generalized zeroth laws).

Now, let us assume that the particular field configuration under consideration
admits a set of covariantly constant YM parameters on $\mathcal{BH}$ that we
label with an index $I$, $\kappa_{I}{}^{A}$

\begin{equation}
  \mathcal{D}\kappa^{A}_{I} \stackrel{\mathcal{BH}}{=} 0\, ,
  \,\,\,\,\,\,
  \Rightarrow
  \,\,\,\,\,\,
  P_{k}{}^{A} \stackrel{\mathcal{BH}}{=} \Phi^{I}\kappa^{A}_{I}\, ,
\end{equation}

\noindent
where the constants $\Phi^{I}$ will be interpreted as the potentials
associated to the YM charges $\mathcal{Q}_{I}$ computed with the parameter
$\kappa_{I}{}^{A}$ Eq.~(\ref{eq:YMcharge})

\begin{equation}
  \mathcal{Q}_{I}
  \equiv
  \mathcal{Q}[\kappa_{I}]
  =
\frac{g_{s}^{(d)\, 2}}{16\pi G_{N}^{(d)}}\int_{\mathcal{BH}}
(-1)^{d-1}\frac{\alpha'}{2}\left[
  e^{-2\phi}\star d\Psi_{I}
  +(-1)^{d}\Psi_{I}\wedge\left(e^{-2\phi}\star H^{(0)}\right)\right]\, ,
\end{equation}

\noindent
where

\begin{equation}
   d\Psi_{I} = -\kappa_{I\, A}F^{A}\, .
\end{equation}

As a result, the third line in Eq.~(\ref{eq:deltaintQ1-2}) becomes
$\Phi^{I}\delta \mathcal{Q}_{I}$.

Now, following Refs.~\cite{Copsey:2005se,Compere:2007vx}, as a consequence of
the KR restricted generalized zeroth law
Eq.~(\ref{eq:KRrestrictedgeneralizedzerothlaw}), we can write (Hodge
decomposition)

\begin{equation}
  P_{k}  -\frac{\alpha'}{2}\left(\Psi_{P_{k}}+\Pi_{P_{k}}\right)
  \stackrel{\mathcal{BH}}{=}
  d e + \Phi^{i}\Lambda_{h\, i}\, ,
\end{equation}

\noindent
where $e$ is some function, the $\Lambda_{h\, i}$ are the harmonic 1-forms of
the bifurcation sphere and the $\Phi^{i}$ are constants that can be
interpreted as the potentials associated to the KR charges
$\mathcal{Q}_{i}=\mathcal{Q}(\Lambda_{h\, i})$ Eq.~(\ref{eq:QLambdacharge})

\begin{equation}
  \label{eq:QLambdachargei}
  \mathcal{Q}_{i}
  =
  -\frac{g^{(d)\, 2}_{s}}{16\pi G_{N}^{(d)}}\int_{C_{\Lambda_{h\, i}}}e^{-2\phi}\star H\, ,
\end{equation}

\noindent
where $C_{\Lambda_{h\, i}}$ is the $(d-3)$-cycle dual to the harmonic 1-form
$\Lambda_{h\, i}$ in $\mathcal{BH}$.

As a result, the fourth line in Eq.~(\ref{eq:deltaintQ1-2}) becomes
$\Phi^{i}\delta \mathcal{Q}_{i}$ and we are left with the first two, which are
linear in the Lorentz momentum map $P_{k}{}^{ab}$, which, on $\mathcal{BH}$,
is given by $\kappa n^{ab}$, where $n^{ab}$ is the binormal to the
horizon. The terms in those two lines must, therefore, be interpreted as those
giving rise to the term $T\delta S$ in the first law

\begin{equation}
\delta M = T\delta S +\Phi^{I}\delta\mathcal{Q}_{I}
+\Phi^{i}\delta\mathcal{Q}_{i} +\Omega^{n}\delta J_{n}\, .
\end{equation}

\section{Wald entropy}
\label{sec-entropy}

It follows from the results of the previous section that the entropy is given
by

\begin{equation}
  \label{eq:Waldentropyformula}
  S
  =
   (-1)^{d}\frac{g_{s}^{(d)\, 2}}{8G_{N}^{(d)}}
 \int_{\mathcal{BH}}
 e^{-2\phi}
 \left\{
   \left[
     \star (e^{a}\wedge e^{b})
        +\frac{\alpha'}{2}e^{-2\phi}\star R^{(0)}_{(-)}{}^{ab}
    \right]n_{ab}
          +(-1)^{d}\frac{\alpha'}{2}\Pi_{n}\wedge \star H^{(0)}
        \right\}\, ,
\end{equation}

\noindent
where we have the defined the 1-form $\Pi_{n}$ (vertical Lorentz
momentum map associated to the binormal) on the bifurcation sphere

\begin{equation}
  d\Pi_{n}
  \stackrel{\mathcal{BH}}{=}
  R^{(0)}_{(-)}{}^{ab}n_{ab}\, .
\end{equation}

This is the main result of this paper, which we will discuss in the next
section. It is worth stressing that the term that involves $\Pi_{n}$, and
which has been shown to given an important contribution to the entropy of
well-known black-hole solutions
Refs.~\cite{Cano:2018qev,Cano:2018brq,Cano:2019ycn,Elgood:2020xwu,Ortin:2020xdm}
occurs in the entropy formula just to cancel an equivalent term that we had to
add to get the correct definition of the KR charge and the associated
potential. Without a detailed knowledge of the conserved charges, the
restricted generalized zeroth laws and the potentials associated, the
presence of that term in the entropy formula could not have been guessed.

\section{Discussion}
\label{sec-discussion}

In this paper we have derived an entropy formula for the black-hole
solutions of the Heterotic Superstring effective action to first order
in $\alpha'$ using Wald's formalism \cite{Lee:1990nz,Wald:1993nt}
taking carefully into account all the symmetries of the theory. A a
result, our entropy formula Eq.~(\ref{eq:Waldentropyformula}) is
manifestly gauge invariant. In particular, it is manifestly invariant
under local Lorentz transformations.

It is interesting to compare this result with the one that would
follow form the direct (and naive) application of the Iyer-Wald
prescription \cite{Iyer:1994ys}. The first two terms in
Eq.~(\ref{eq:Waldentropyformula}) can be obtained from
Eq.~(\ref{eq:heterotic1diff}) by varying the Einstein-Hilbert term and
the $R^{2}_{(-)}$ term with respect to the Riemann curvature tensor,
but the third term cannot be obtained in that way from the $H^{2}$
term. As stressed in
Refs.~\cite{Cano:2019ycn,Elgood:2020xwu,Ortin:2020xdm}, the variation
of this term with respect to the Riemann tensor gives a term of the
form

\begin{equation}
  \label{eq:thatterm}
  \frac{\alpha'}{4} e^{-2\phi} \left(\Omega^{(0)}_{(-)}{}^{ab}n_{ab}\right)
  \wedge \star H^{(0)}\, ,   
\end{equation}

\noindent
which is not Lorentz-covariant. The coefficient of this term differs
from the last term in Eq.~(\ref{eq:Waldentropyformula}) if we
associate $\Pi_{n}$ to $\Omega^{(0)}_{(-)}{}^{ab}n_{ab}$, which is the
right thing to do as we are going to show. But this coefficient
changes after dimensional reduction, as observed in
Ref.~\cite{Faedo:2019xii}. The explicit calculation in
Ref.~\cite{Cano:2019ycn} shows that the right coefficient is the one
that arises after dimensional reduction,\footnote{The entropy
  calculated in this way satisfies the first law or, equivalently, the
  thermodynamic relation
  \begin{equation}
  \frac{\partial S}{\partial M} = \frac{1}{T}\, .  
  \end{equation}
}
but, certainly, there are ambiguities in the way in which the Chern-Simons
terms are defined in lower dimensions.

It is interesting to observe that because
$\mathcal{D}n_{ab}\stackrel{\mathcal{BH}}{=}0$,

\begin{equation}
  d\Pi_{n}
  \stackrel{\mathcal{BH}}{=}
d\left(\Omega^{(0)}_{(-)}{}^{ab}n_{ab}\right)
+\Omega^{(0)}_{(-)}{}^{a}{}_{c}\wedge\Omega^{(0)}_{(-)}{}^{cb}n_{ab}\, .
\end{equation}

For the non-extremal Reissner-Nordstr\"om black hole of
Ref.~\cite{Khuri:1995xq}, whose $\alpha'$ corrections were computed in
Ref.~\cite{Cano:2019ycn}, the second term vanishes identically in the
tangent space basis used (see Appendix~C). This shows that, in that
basis, our entropy formula and the entropy formula obtained via the
Iyer-Wald prescription (after dimensional reduction) give the same
result. Of course, our formula is valid in any basis.

Our entropy formula seems to differ from the entropy formula obtained
in Ref.~\cite{Edelstein:2019wzg}, but a detailed comparison is not
possible since that formula contains undetermined parameters that
guarantee its invariance under Lorentz transformations. In
Ref.~\cite{Edelstein:2019wzg} it was argued that those undetermined
parameters do not contribute to the entropy in certain cases but,
without an explicit expression, it is difficult to understand why or
when this may happen. Furthermore, as we have shown, the
identification of the entropy formula can only be made after the first
law of black hole mechanics has been proven and this requires a
careful identification of the conserved charges of the theory: some
terms (the one involving $\Pi_{n}$) occur in the entropy formula only
because they are needed to compensate other terms that have to appear
in the correct definition of the KR charge. This analysis was simply
not carried out in Ref.~\cite{Edelstein:2019wzg}.

Our entropy formula (the contribution due to the presence of Lorentz-
or gravitational Chern-Simons terms in $H^{(1)}$) also differs from
the one found in Ref.~\cite{Tachikawa:2006sz}. Observe that Eq.~(40)
in Ref.~\cite{Tachikawa:2006sz}, similar to the terms contains in the
formulae derived in Refs.~\cite{Elgood:2020xwu,Ortin:2020xdm} and to
Eq.~(\ref{eq:thatterm}) is not covariant. Thus, it may give the right
result in certain basis, if at all.\footnote{the non-covariance of
  Tachikawa's entropy formula was observed in
  Ref.~\cite{Azeyanagi:2014sna}, where an alternative method was
  devised to deal with this problem. Nevertheless, the formula
  obtained in Ref.~\cite{Azeyanagi:2014sna} reduces to Tachikawa's in
  $\mathcal{BH}$, apparently losing the covariance, while ours does
  not. } The problems in the derivation of
Ref.~\cite{Tachikawa:2006sz} are having overlooked the KR conserved
charge and the determination of the gauge parameters that generate
symmetries of the complete field configuration.

Finally, it is interesting to notice that the entropy formula looks like the
charge associated to the Lorentz transformations generated by the binormal to
the horizon. These transformations preserve the connections $\omega$ and
$\Omega_{(-)}^{(0)}$ on the bifurcation sphere, but they do not preserve the
Vielbein, as we assumed in Section~\ref{sec-Lorentzcharge}
(Eq.~(\ref{eq:kappae=0})), which produces an additional term associated to the
Einstein-Hilbert term.

The main use of the entropy formula that we have found is to put in
solid ground the calculations of the macroscopic entropies of
$\alpha'$-corrected black holes, an ineluctable condition for a fair
comparison with the microscopic ones. More $\alpha'$-corrected
solutions will be available to this end \cite{kn:CORRZ}. As mentioned
in the introduction, another necessary ingredient for this comparison
is the correct identification of the relation between the charges of
the black hole and the branes in the string background.  These results
and those of our previous work \cite{Elgood:2020mdx} single out a very
precise definition of the conserved charges, which turn out to be of
\textit{Page type}, conserved and gauge-invariant under the
assumptions made. This fact should shed light on this problem and we
intend to pursue this line of research in future work.

\section*{Acknowledgments}

TO would like to thank G.~Barnich, P.~Cano, P.~Meessen,
P.F.~Ram\'{\i}rez, A.~Ruip\'erez and C.~Shahbazi for many useful
conversations and their long-term collaboration in this research
topic. DP would also like to thank G.~Barnich for many useful
conversations. This work has been supported in part by the MCIU, AEI,
FEDER (UE) grant PGC2018-095205-B-I00 and by the Spanish Research
Agency (Agencia Estatal de Investigaci\'on) through the grant IFT
Centro de Excelencia Severo Ochoa SEV-2016-0597.  TO wishes to thank
M.M.~Fern\'andez for her permanent support.

\appendix


\end{document}